\documentclass[numberedappendix,appendixfloats,iop]{emulateapj}
\usepackage{natbib}
\usepackage[utf8x]{inputenc}
\usepackage{amssymb}
\usepackage{amsmath}
\usepackage{graphicx}
\usepackage{setspace}
\usepackage{color}
\def \rec {\text{rec}}
\def \cl {\text{cl}}

\def \out {\text{out}}
\def \bb {\text{BB}}

\def \out {\text{out}}
\def \delt { \left( \frac{m_{\cl}(t_{\rec})}{\hat{m}(t_{\rec})} \right) }

\def \ad {\text{ad}}
\def \in {\text{in}}
\def \ph {\text{ph}}
\def \sn {\text{SN}}
\def \ion {\text{ion}}
\def \erase #1 {}

\begin{document}

\title{Recombination Effects on Supernovae Light-Curves}
\author{Tomer Goldfriend\altaffilmark{1,2}, Ehud Nakar\altaffilmark{2} and Re'em Sari\altaffilmark{1}}
\altaffiltext{1}{Racah Institute for Physics, The Hebrew University, Jerusalem, 91904, Israel}
\altaffiltext{2}{The Raymond and Berverly Sackler School of Physics and Astronomy,
Tel Aviv University, 69978 Tel Aviv, Israel}

\email{goldfriend@poat.tau.ac.il}

\begin{abstract}
Supernovae of type  IIP are marked  by the long  plateau seen  in their
optical  light  curves.  The  plateau  is  believed to  be  the  result  of  a
recombination  wave that  propagates through  the outflowing  massive hydrogen
envelope.  Here,  we analytically  investigate  the  transition from  a  fully
ionized envelope to a partially recombined one and its effects on the SN light
curve. The motivation is to  establish the underlying processes which dominate
the evolution  at late times when  recombination takes place in  the envelope,
yet early enough so that $^{56}$Ni decay  is a negligible source of energy. We
assume a simple, yet adequate, hydrodynamic  profile of the envelope and study
the  mechanisms   which  dominate  the   energy  emission  and   the  observed
temperature. We consider  the diffusion of photons through  the envelope while
analyzing the ionization fraction and  the coupling between radiation and gas.
We find  that once  recombination starts,  the observed  temperature decreases
slowly in  time. However,  in a  typical red  supergiant (RSG)  explosion, the
recombination wave does not affect the bolometric luminosity immediately. Only
at later  times, the  cooling wave may  reach layers that  are deep  enough to
affect the luminosity.  We find that the plateau is not a generic  result of a
recombination process  in expanding gas. Instead it depends  on the
density profile of the parts of  the envelope which undergo recombination. Our
results are useful to investigate the  light curves of RSG explosions. We show
the resulting light curves of two  examples of RSG explosions according to our
model  and discuss  their  compatibility with  observations.  In addition,  we
improve the  analytical relations between  the plateau luminosity  and plateau
duration    to    the    properties   of    the    pre-explosion    progenitor
\cite[]{arnett80,popov93}.

\end{abstract}


\section{Introduction}

Type IIP supernova (SN) is the  most common supernova type \citep{Li11}. It is
marked by  H lines  in its spectra  and by a  long, $\approx$100  days optical
plateau that starts between a few days to a few weeks after the explosion. The
progenitors  of type  IIp SNe  are red  supergiants (RSGs),  which exhibit  an
extended   H   envelope   \cite[]{smartt04,van_dyk03,maund09}.   The   initial
temperature of the envelope as it  starts to expand, after the shock crossing,
is high  ($> 10^5$  K), but  it drops with  time due  to adiabatic  loses. The
observed  temperature  drops rapidly  until  it  reaches the  H  recombination
temperature,  $\approx$7000 K,  where it  remains rather  constant during  the
entire plateau phase. For that reason it is believed that recombination is the
source       of       the       long      lasting       optical       plateau.

The initial conditions for the beginning of the recombination phase are set by
the radiation dominated shock that crosses the envelope and explodes the star.
The shock accelerates in the decreasing  density gradient near the edge of the
stellar envelope  and determines the  velocity and internal energy  profile of
the envelope. Once the shock reaches to a point where the distance to the edge
of the star is comparable to the shock's width, it ``breaks out''. This is the
first      electro-magnetic      signal       of      a      SN      explosion
\cite[]{colgate74,folk78,klein_chevalier78,imshennik_etal81}.
The envelope  continues  to  accelerate  while photons  continuously  diffuse  out
through it \cite[]{matzner&mckee}. The acceleration ends before a considerable
expansion takes place  and at the end  of this phase there is  a hot, ionized,
radiation dominated envelope  in a homologous expansion.  The internal regions
are dense  with large optical  depth and  they lose energy  adiabatically. The
external regions  have a lower optical  depth and photons can  diffuse through
them over a  dynamical time. The expansion reduces the  envelope optical depth,
enabling photons to diffuse from deeper  regions in the envelope. At breakout,
the temperature  of the external regions  from where radiation can  diffuse to
the observer is $\sim 30$ eV  \cite[]{ns10}. It drops over a week or
two until it reaches the recombination temperature. Until that time the entire
envelope is  ionized and  its opacity is  Thomson opacity.  Once recombination
starts in the  outer layers it leads to  a sharp drop in the  opacity of these
layers. The  fast escape of radiation  via the low opacity  recombined regions
causes a fast drop  in the temperature also in the  ionized regions which lies
just  behind the recombined  layer. This  cooling leads  to farther
recombination  which in  turn leads  to drop  in the  opacity and  so on.  The
resulting  picture  is  a  recombination  wave,  followed  by  cooling,  which
propagates inward (in Lagrangian sense)  through the envelope. The propagation
of this wave  depends on the velocity  and on the internal  energy profiles of
the envelope  and obtaining an analytic  understanding of this process  is the
goal                     of                    this                     paper.   

In  this  work  we  derive  an   analytical  model  to  study  the  effect  of
recombination on  SNe light  curves. We  examine the  problem of  a homologous
expanding envelope  which cools adiabatically, together  with energy transport
by  diffusion of  photons. We  assume that  $^{56}$Ni radioactive  decay is  a
negligible source of energy. This condition holds for a typical explosion of an
RSG, in which the radioactive decay  luminosity is low compared to the release
of the thermal energy deposited by the shock (see discussion in appendix \ref{app:Ni}).
In addition, we neglect the effect of radioactive decay on the opacity, due to
non-thermal   photoionization.   Our   model   is   based   on   recent   work
by~\citet[]{ns10}   (NS10  hereafter)   where  early   light  curves,   before
recombination has a  significant role, were derived. As a  result, we can draw
the evolution of  the light curves from the initial  pulse (shock breakout) up
to the time when most of the  envelope is recombined. We find that
it  is  essential  to  take  into account  the  coupling  between  matter  and
radiation since in the recombined gas the opacity  can vary by
orders of magnitude while the temperature changes only by a factor of two. For
that reason it  is essential to find  the exact temperature in  this layer and
thus the  opacity drop. Here, the radiation-gas coupling plays  the main
role since  a drop  in the  temperature reduces the  coupling (along  with the
opacity), while  in order for  the temperature to  drop a minimal  coupling is
required. This  feedback of the  coupling on the  temperature is what  sets the
exact temperature  and opacity drop in  the recombined layer. By  solving self
consistently for the  radiation-gas coupling and the opacity  drop, within the
expanding gas  velocity and internal energy  profiles, we find a  solution for
the  propagation of  the recombination  wave  and for  the resulting  observed
radiation.

The  propagation  of  a  recombination   wave  in  SNe  envelope  was  studied
analytically                 in                 previous                 works
\citep{grassberg71,grass_nad76,popov93,rabinak11}.  However, these  works used
more limited assumptions compared to our  model, such as full coupling between
gas and radiation. In addition, some of these models were derived for physical
conditions  which are  not applicable  for SNe  envelope, such  as homogeneous
envelope \cite[]{popov93}. A  comparison to these previous  works is discussed
in  section \ref{sec:comparison}.  We also  compare qualitatively  between our
results   and   the   results   from   recent   numerical   works,   such   as
\cite{utrobin07,woosley&kasen09}  and  \cite{bersten11},  which  examined  the
light  curves of  type IIPs  and  discussed in  detail the  properties of  the
recombination            wave            in           the            envelope.

The paper  is organized as  follows. In section \ref{sec:shells}  we introduce
the hydrodynamical  profile of the  envelope we use  in our model.  In section
\ref{sec:NS} we  discuss the  evolution at  early stages, up  to the  onset of
recombination,  according to  NS10.  Then, in  section \ref{sec:Receffect}  we
describe how recombination,  through the decrease in  the ionization fraction,
affects the diffusion of energy and production of new photons in the envelope.
In  section \ref{sec:LC}  we introduce  the  solution to  these equations  and
calculate  the   luminosity  and   the  observed   temperature  for   a  given
hydrodynamical  profile of  the envelope.  We treat  separately two  different
cases.  The first  is  the realistic  case where  the  ionization fraction  is
parametrized as a continuous function of the temperature (\S\ref{sec:mrec}). The
second is the academic, yet enlightening, case where opacity is parametrized by a
step  function  of  the  temperature (appendix  \ref{app:const}).  In  section
\ref{sec:application} we apply the  results of \S\ref{sec:mrec} and
appendix \ref{app:const}  to an analytic description of a typical RSG expanding envelope. In addition,  we apply
the model to  profiles given numerically from a simulation of  RSG explosion and
derive   the  corresponding   light  curves   semi-analytically.  In   section
\ref{sec:comparison} we compare results from previous analytical and numerical
works to our  results. Finally in \ref{sec:summary} we  summarize our findings.

\section{The  Hydrodynamic  Profile  of   the  Envelope  }  \label{sec:shells}

The expanding  envelope can  be considered  as a  series of  successive shells
(NS10).  For   each  shell,   there  are  no   considerable  changes   in  the
hydrodynamical parameters  over the shell  width. Following the  breakout, the
evolution of each  shell can be divided into two  phases: planar phase, before
the shell  radius doubles, and  spherical phase at  later times. Since  we are
interested in  later times we discuss  only the spherical phase  of the shells
which  takes place  after  acceleration  ends and  the  envelope expansion  is
homologous.  We treat the density  and adiabatic  energy
profiles of the envelope during the  spherical phase as given. We parametrize
these profiles  by power laws.  The applicability of such  parametrization for
SNe             envelope            is             discussed            below.

The evolution of the density is given by
\begin{equation}
\rho(r,t)=f_{\rho} r^{-k} t^{k-3} ,
\label{eq:rhoprof}
\end{equation}
where $k$ is a positive constant and $f_{\rho}$ is a constant 
which depends on the initial properties of the progenitor.
Each mass shell is characterized by its mass, $m$.
During the spherical phase, the width of each shell is comparable to the radius of the shell
$r \approx v(m)t$, thus, equation \eqref{eq:rhoprof} corresponds to the velocity profile
\begin{equation}
v(m) \propto m^{-\frac{1}{k-3}} .
\label{eq:vprof}
\end{equation}
At a given time we parametrize the internal energy in the regions that cools adiabatically as
\begin{equation}
E_{\ad}(m,t)=f_{\ad} m^{s} r^{-1} ,
\label{eq:eprof}
\end{equation}
where $s$ is a positive constant and $f_{\ad}$ depends on the initial conditions of the progenitor and the explosion.
$E_{\ad}$ depends on $t$ through $r=v(m)t$.

Equation \eqref{eq:vprof} shows that the parameter $k$ is related to the coasting velocity profile of the envelope.
In order to determine the parameter $s$, additional information about 
the profiles of the velocity and density right after the passage of the shock, $v_{i}(m)$ and $\rho_i(m)$, is needed.
This is because the initial thermal energy of each shell following the breakout is $\sim mv_i^2$,  so 
$E_{\ad}(m,t)=m v_i^2 (\rho/\rho_i)^{1/3}=m v_i^2 (m/\rho_i)^{1/3} r^{-1}$. 

The power laws parametrization is applicable at the external part of the SN envelope, located 
internal to the
point from which the shock breaks out ($\sim 10\%$ of the envelope mass).
This part controls the light curves at early stages.
The profiles in these external parts can be approximated analytically with $k=9.5-12$ and $s=0.8-0.9$ \citep[][ and references therein]{matzner&mckee}.
At late stages of the evolution, deeper shells become transparent and control the light curve.
These shells are characterized with softer density profiles,
and a description of the evolution using power-laws is less accurate.

In sections \ref{sec:NS} and \ref{sec:LC}
we provide a general solution for single power law profiles described by the parameters $k$ and $s$.
For simplicity, together with the general solution, we introduce our results for specific values of $k$ and $s$. 
For the outer layers (section \ref{sec:NS}) we use $k=12$ and $s=0.9$.
For inner parts, where recombination takes place  (section \ref{sec:LC}), we take representative values that are deduced from a numerical simulation.
The basic equations of our model (introduced in \S\ref{sec:rec_shell}), which govern the evolution,
can be solved numerically for any given profiles of the envelope.   
Example of a numerical solution is given in
\S\ref{sec:example}.

\section{Early Light-Curves} \label{sec:NS}

\subsection{Luminosity Shell and Color Shell}

While the envelope is highly ionized and all the relevant shells are at their spherical phase,
the optical depth and the diffusion time of each shell are
\begin{equation}
\tau \approx \kappa_{T} \rho r
\quad , \quad
t_{d} \approx \frac{\tau r}{c} ,
 \label{eq:tau_nr}
\end{equation}
where $\kappa_T$ is Thomson opacity for ionized gas and $c$ is the speed of light.
The luminosity is dictated by the luminosity shell from which photons diffuse out effectively.
In this shell the diffusion time is equal to the time since explosion, which is also the dynamical time
of the shell- $r/v(m)$ .
Photons in shells internal to the luminosity shell can barely escape because the diffusion time there is longer.
On the other hand, shells external to the luminosity shell have already
released their energy at earlier times.  
At any given time
we can find the mass of the luminosity shell by requiring 
\begin{equation}
\tau(\hat{m},t)=c/v(\hat{m}) .
\label{eq:old_lum_shell}
\end{equation}
The properties of the luminosity shell are marked with superscript$\hat{\quad}$.
Here we repeat the analysis of NS10 in terms of the parameters $k$ and $s$. Using equations \eqref{eq:vprof}, \eqref{eq:eprof} and
\eqref{eq:tau_nr}, and the definition of the luminosity shell \eqref{eq:old_lum_shell}, we find
$$\hat{m}(t) \propto t^{\frac{2(k-3)}{k-2}} \approx t^{1.8} ,$$
\begin{equation}
 L(t)=\hat{E}_{\ad}/t \propto t^{-\frac{2(1-s)(k-3)}{k-2}} \approx t^{-0.17} .
 \label{eq:lum_old}
\end{equation}
In this section we use  $k=12$ and $s=0.9$ as canonical values.

The energy density in shells internal to the luminosity shell is dominated by adiabatic cooling 
(equation \eqref{eq:eprof}). 
At any given time the luminosity through shells external to the luminosity shell is
constant and given by equation~\eqref{eq:lum_old}.
The energy density, $\epsilon$, in these shells is determined by the diffusion of total luminosity $L$ with
a diffusion time $t_d$ given in equation \eqref{eq:tau_nr},
\begin{equation} \label{eq:eps}
   \epsilon(m,t)= \left\{
  \begin{array}{l l}
   \frac{E_{\ad}}{r^3} & \quad   m>\hat{m}\\
    \\
   \frac{L \tau}{cr^2} & \quad   m<\hat{m}\\
  \end{array} \right. .
\end{equation}
Note that the above expression holds up to the shell in which $\tau \approx 1$.

When the radiation is thermalized, the temperature is a blackbody temperature given by
\begin{equation}
 T_{\bb}=(\epsilon/a)^{1/4} .
\label{eq:Tbb}
\end{equation} Where $a$ is the radiation constant.
The optical depth of the luminosity shell is always greater than 1. Thus, photons which escape from the luminosity shell
can have many interactions with the gas as they are traveling through the envelope until they finally leave it  
at the point where the optical depth is $\sim 1$.
For the relevant physical conditions the luminosity shell is in thermal equilibrium ($\hat{T}=\hat{T}_{\bb}$), see \S\ref{sec:dawn}.

Consider shells at the region external to the luminosity shell - $r>\hat{r}$.
The photons dominate the heat capacity, hence, 
the total radiation flux in this region is independent of the coupling between the photons and the gas.
As long as the radiation is in thermal equilibrium, the photon number flux increases with $r$
and the typical energy of each photon, which originated in the luminosity shell, 
is changed while it travels outward.
Thus, in order to keep the radiation thermalized, electrons in each shell must generate sufficient amount
of new photons which share their energy with photons which arrive from internal shells.
External to the point in which the radiation departs from thermal equilibrium, the photon number flux is fixed 
(at a given time) and the temperature can not change.
Hence, the observed temperature is the blackbody temperature of the outermost shell which is in
thermal equilibrium.

We  assume  that  free-free  emission  is the  dominant  process  for  photons
production.  Similarly  to  NS10  we define  a  thermal  coupling  coefficient
\begin{equation}  \label{eq:eta_def} \eta  \equiv \frac{n_{\text{BB}}}{  t_{d}
\cdot  \dot{n}_{\text{ph,ff}}(T_{\text{BB}})}  \quad  ,  \end{equation}  where
$n_{\text{BB}}=aT_{\bb}^3/3k_B$ is the density of  photons with $h \nu \approx
3k_BT_{\text{BB}}$ and  $\dot{n}_{\text{ph,ff}}(T)= 3.5 \times  10^{36} \text{
s}^{-1} \text{cm}^{-3} \rho^{2} T^{-1/2}$ is  the free-free emission rate, per
unit volume, of photons with energy $h\nu \approx 3k_BT$. The parameter $\eta$
defined this way, is the time required to achieve thermal equilibrium ignoring
photon  diffusion $\sim  n_{\text{BB}}/\dot{n}_{\text{ph,ff}}(T_{\text{BB}})$,
divided  by the  diffusion time.  Photons released  from the  luminosity shell
($\hat{\eta}<1$)  change  their  typical  energy  up to  the  shell  in  which
$\eta=1$, where  the observed  temperature is  determined. Thus,  the observed
spectrum is blackbody at a temperature of that shell, which we call 
the ``color"  shell. We  denote the  properties of the  color shell  with the
subscript $_{\cl}$. The evolution of the color temperature and the color shell
mass are given by  
$$T_{\cl}(t) \propto t^{\frac{2(k-3)(6ks-11k-4s+14)}{(k-2)(17k-23)}}\approx t^{-0.56} , $$
\begin{equation}
m_{\cl}(t) \propto t^{-\frac{14(-ks-k+3s+1)(k-3)}{(k-2)(17k-23)}} \approx t^{1.33},
\label{eq:dawn_cl}
\end{equation}
where the color shell is defined by $\eta(m_{\cl},t)=1$. The gas and radiation must have one temperature up to the color shell, but outward of the color shell they can decouple, where the radiation temperature is constant while the gas temperature may drop.   The complete equations of the color shell evolution are (4),(6),(19) and (21)  in NS10. 
The pre-factors are related to the initial conditions of the progenitor.
For further discussion on the coupling between radiation and matter in ionized medium see NS10.

\subsection{Initial conditions for recombination} \label{sec:dawn}

We summarize the state of the envelope just before recombination starts.
That is, we draw the initial conditions for the problem we solve in the next sections. 
We assume that just before recombination starts the envelope is in its spherical phase, i.e. all the shells which
dominate the light curves are in their spherical phase. 
In addition, we assume that the luminosity shell is already in thermal equilibrium ($\hat{\eta}<1$) so the color shell
is located external to the luminosity shell.
These physical conditions are adequate for an explosion of a RSG (NS10).

\begin{figure}
    \includegraphics[width=0.5\textwidth]{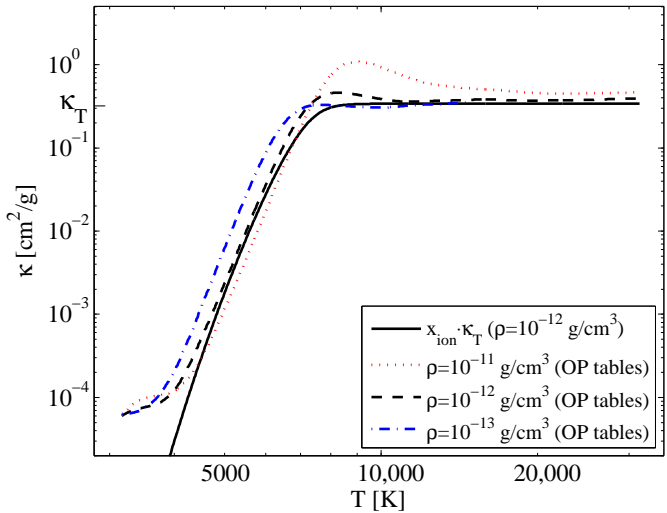}
    \caption{  Rosseland opacity using OP tables~\cite[]{seaton2005} and Saha equation. The dash-dotted lines
    are Rosseland opacities for Y=0.3, Z=0.02 (solar metalicity) taken from the tables. The solid line is a theoretical line
    $\kappa=x_{\text{ion}} \kappa_{_{T}}$ where $x_{\text{ion}}$ is the ionization fraction given by Saha equation. }
\label{fig:ross}%
\end{figure}

Before recombination starts the evolution of the luminosity shell and the luminosity
are given in equation \eqref{eq:lum_old}. 
\textit{The recombination temperature}, $T_{\rec}$, is defined as the temperature in which there is a sharp change in
the opacity~(figure \ref{fig:ross}).
When the color temperature 
drops below
the recombination temperature
we must take into account the effect of recombination.
We define the recombination time, $t_{\rec}$, by 
$T_{\text{cl}}(t_{\rec})=T_{\rec}$.
In the next section we find the evolution of the luminosity shell and the color shell after the onset of 
recombination. The solution is written as a function of the conditions when recombination starts: 
$\hat{m}(t_{\rec})$, $m_{\cl}(t_{\rec})$, $L(t_{\rec})$ and $T_{\cl}(t_{\rec})=T_{\rec}$.  

\section{The effect of recombination} \label{sec:Receffect}
\subsection{Modeling the Opacity and the Free-Free Process}

Recombination changes the opacity and the rate of generation of new photons.
In section~\ref{sec:NS}, considering the evolution before recombination starts, 
we take these quantities to be $\kappa=\kappa_T$ and
 $\dot{n}_{\text{ph}} = \dot{n}_{\text{ph,ff}} = 3.5 \times 10^{36} \rho^2 T^{-0.5}$
respectively. In our model, we assume that when recombination occurs,
electron scattering still dominates the opacity and that free-free emission still dominates the photon production rate\footnote{Even
in the case that bound-free is the dominant process, which might be the case
when recombination occurs, the expression for the production rate differs from that of the free-free process only by a constant that depends (linearly) on the metalicity.}.
Thus, we write
$\kappa=x_{\ion} \cdot \kappa_{T}$ and 
$\dot{n}_{\text{ph}} =x_{\ion}^2 \dot{n}_{\text{ph,ff}} $,
where $x_{\ion}$ is the ionization fraction. These two quantities decrease as electrons and
ions become less abundant.
Figure~\ref{fig:ross} shows that for the low densities ($\le 10^{-12} \text{g cm}^{-3}$) and temperatures between 3000 K to 12,000 K
, indeed Thomson scattering provides almost all the opacity. This range of densities is the one found in RSG explosions at $t_{rec}$.
A good approximation for the ionization fraction can be obtained from the Saha equation.
However, to obtain analytical results, we parametrize the ionization fraction as
\begin{equation} \label{eq:alpha}
   x_{\ion} = \left\{
  \begin{array}{l l}
    1 & \quad   T>T_{\rec}\\
    \left( T/T_{\rec}\right)^{11}  & \quad T<T_{\rec}\\
  \end{array} \right. .
\end{equation}
In addition, in appendix~\ref{app:const} we derive an analytical solution 
for the parametrization of the ionization fraction
as a step function, equation \eqref{eq:alpha_const}.
 
We neglect the effects of velocity gradients on opacity through atomic line broadening
and the increase of ionization ratio by non-thermal $\gamma$ rays from $^{56}$Ni radioactive decay (see discussion in section \ref{sec:summary}).
We also neglect the energy released by recombination (see section \ref{sec:comparison}).

\subsection{Basic Equations} \label{sec:rec_shell}

Recombination starts to affect the
evolution when the color temperature equals the recombination temperature.
Let us consider the envelope from the luminosity shell out to the point where $T=T_{\rec}$.
The energy density of this part of the envelope is dictated by the diffusion of photons (equation \eqref{eq:eps}).
At the point where $T=T_{\rec}$ the opacity drops and photons escape ``more easily'' causing a sharp decrease in the
energy density. This drop is related to a drop in the temperature which in turn decreases the opacity.
Therefore, from $t=t_{\rec}$ recombination moves inward (in the Lagrangian sense) and starts to reach deeper shells.
We call the shell in which $T=T_{\rec}$ the \textit{recombination shell}, and denote its properties by the subscript $_{\text{rec}}$.
Because the change in the opacity at $T_{\rec}$ is sharp enough, the energy density and the optical depth profiles changes 
significantly within this shell. 
While these profiles inside the recombination shell are complicated (see, for example figure \ref{fig:mrec_out}), 
the properties of shells located far from the recombination shell (in or out) can be easily determined.
This is because within those shells all the hydrodynamical parameters 
can be approximated as homogeneous.
The optical depth and thermal coupling coefficient of such shells are given by
\begin{equation} \label{eq:tau_r}
\tau(m \neq m_{\rec})=\frac{c}{\hat{v}(t_{\rec})}
 \left(\frac{\rho}{\hat{\rho}(t_{\rec})} \right)
\left(\frac{r}{\hat{r}(t_{\rec})} \right)
x_{\ion}(T) ,
\end{equation}
\begin{equation}
 \eta(m \neq m_{\rec})= \left(\frac{T_{\bb}}{T_{\rec}} \right)^{\frac{7}{2}}
\left(\frac{\rho}{\rho_{\cl}(t_{\rec})} \right)^{-3} \times
\label{eq:eta_r}
\end{equation}
$$
\left(\frac{r}{r_{\cl}(t_{\rec})} \right)^{-2}
x_{\ion}(T)^{-3} .
$$

It is only the region between the luminosity shell and the color shell, $\hat{r} \leq r \leq r_{\cl}$, which is relevant
in order to understand the evolution, since beyond these two regions both the luminosity and the temperature are constant.
Now, if the dependence of the opacity on temperature is strong enough, the color shell and recombination shell
are the same, i.e., $r_{rec} \cong r_{cl}$, at any time after $t_{rec}$. One can see that by assuming the contrary, that $r_{rec}  \ll r_{cl}$, then  equations (\ref{eq:eps})-(\ref{eq:Tbb}) imply that the temperature is an increasing function of radius ( $T \propto (r/\rho)^{1/7}$ for $x_{\ion} \propto T^{11}$), which is not physical.
Therefore, within the recombination shell the temperature drops sharply and by the end of the shell the system must leave thermal 
equilibrium with $\eta=1$. The radiation temperature is then fixed external of this point, and is not an increasing function of $r$.
This shows that recombination and thermal coupling are intimately related and there could be no
correct treatment of recombinations that ignores thermal coupling.

While the recombination wave is described by the non uniform temperature profile in the recombination shell, we find the dynamics of the recombination shell without dealing with the 
profiles inside it. Instead we consider the properties on its boundaries,
where all the properties - $\epsilon$, $T_{\bb}$, $x_{\ion}$, $\tau$ and $\eta$, 
are approximately homogeneous and given by equations
\eqref{eq:eps}, \eqref{eq:Tbb}, \eqref{eq:alpha}, \eqref{eq:tau_r} and \eqref{eq:eta_r}. We indicate the properties of the internal and external boundary with the sub-scripts
$_{\rec-\in}$ and $_{\rec-\out}$. The internal boundary is located at a distance $\sim r_{\rec}$ inward from
the point where $T=T_{\rec}$ and the external boundary, which coincides with the color shell, at a distance $\sim r_{\rec}$ outward from
that point.
The density of the recombination shell is roughly uniform (does not change on scale much smaller than $r$ like the temperature
or energy density) and is given by $\rho_{\rec} \sim m_{\rec}/(v(m_{\rec}) t)^3$. 

The internal boundary of the recombination shell is not affected by recombination.
The value of $\epsilon_{\rec-\in}$ and $\eta_{\rec-\in}$ can be approximated by equations \eqref{eq:eps}-\eqref{eq:Tbb} and
\eqref{eq:tau_r}-\eqref{eq:eta_r}
with $m=m_{\rec}$ and $x_{\ion}=1$. This implies that
$\eta_{\rec-\in}<1$ and $\epsilon_{\rec-\in}>aT_{\rec}^4$. 
The properties of the external boundary are dictated by the end of thermal coupling - $\eta=1$.
The luminosity in the recombination shell is dictated by the inner region:
the luminosity shell when $\hat{m}>m_{\rec}$ and the inner boundary 
of the recombination shell itself otherwise\footnote{The situation in which $\hat{m}<m_{\rec}$ is
not relevant in our model. The dynamics of $m_{\rec}$ in this case, which is given solely by its adiabatic cooling 
(equation \eqref{eq:eprof}), is inconsistent.}.
A schematic description of the energy density profile within the recombination shell 
is given in figure \ref{fig:mrec_out}.

To summarize, for $t \geq t_{\rec}$, 
the equations governing the dynamics of the three 
characteristic shells are
\begin{subequations}
\begin{equation}
\left \{
  \begin{array}{l l}
   \tau(\hat{m})=c/ v(\hat{m}) & \quad   \text{, if } \hat{m}>m_{\rec}\\
   \hat{m}=m_{\rec} & \quad   \text{, otherwise}\\
  \end{array} \right. ,
\label{eq:num_mhat}
\end{equation}
\begin{equation}
L=\frac{E_{\ad}(\hat{m})}{t} 
=\frac{ aT_{\cl}^4 c r_{\rec} }{x_{\ion}(T_{\cl}) \kappa_T \rho_{\rec} } ,
\label{eq:num_lum}
\end{equation}
\begin{equation}
   \eta_{\rec-\out}=1 .
\label{eq:num_eta}
\end{equation}
\label{eq:general}
\end{subequations}


												
\section{Bolometric Luminosity and Observed Temperature} \label{sec:LC}

We now find the  dynamics of the recombination shell and  the evolution of the
bolometric  luminosity  and color  temperature.  Here  we use  an  approximate
continuous function  of the  ionization fraction  (equation \ref{eq:alpha}),
which closely follows  the Saha equation (figure  \ref{fig:ross}). In appendix
\ref{app:const}  we solve  the  evolution when  the  ionization fraction  is
approximated  by a  step function  (equation \ref{eq:alpha_const}).  A short summary of the main results in this case is presented in \S\ref{sec:const}. The step function model is not realistic and is presented for
deductive purposes and in  order to allows a comparison of  our analysis to an
earlier   model   by   \citet{popov93},  see   section   \ref{sec:comparison}.

We introduce the results in a general form as a function of $k$ and $s$. In addition, we provide
the typical behavior for $k=6$ and $s=0.95$, which are found to be representative based on the numerical simulation presented in the next section.
The solution is written as a function of the state of the envelope just before
recombination. We denote the ratio between the luminosity and color shell masses at the beginning of
recombination by  
$$\Delta \equiv \delt<1.$$

\subsection{Ionization Fraction as a Continuous Function of $T$} \label{sec:mrec}

\begin{figure*}
\includegraphics[width=\columnwidth]{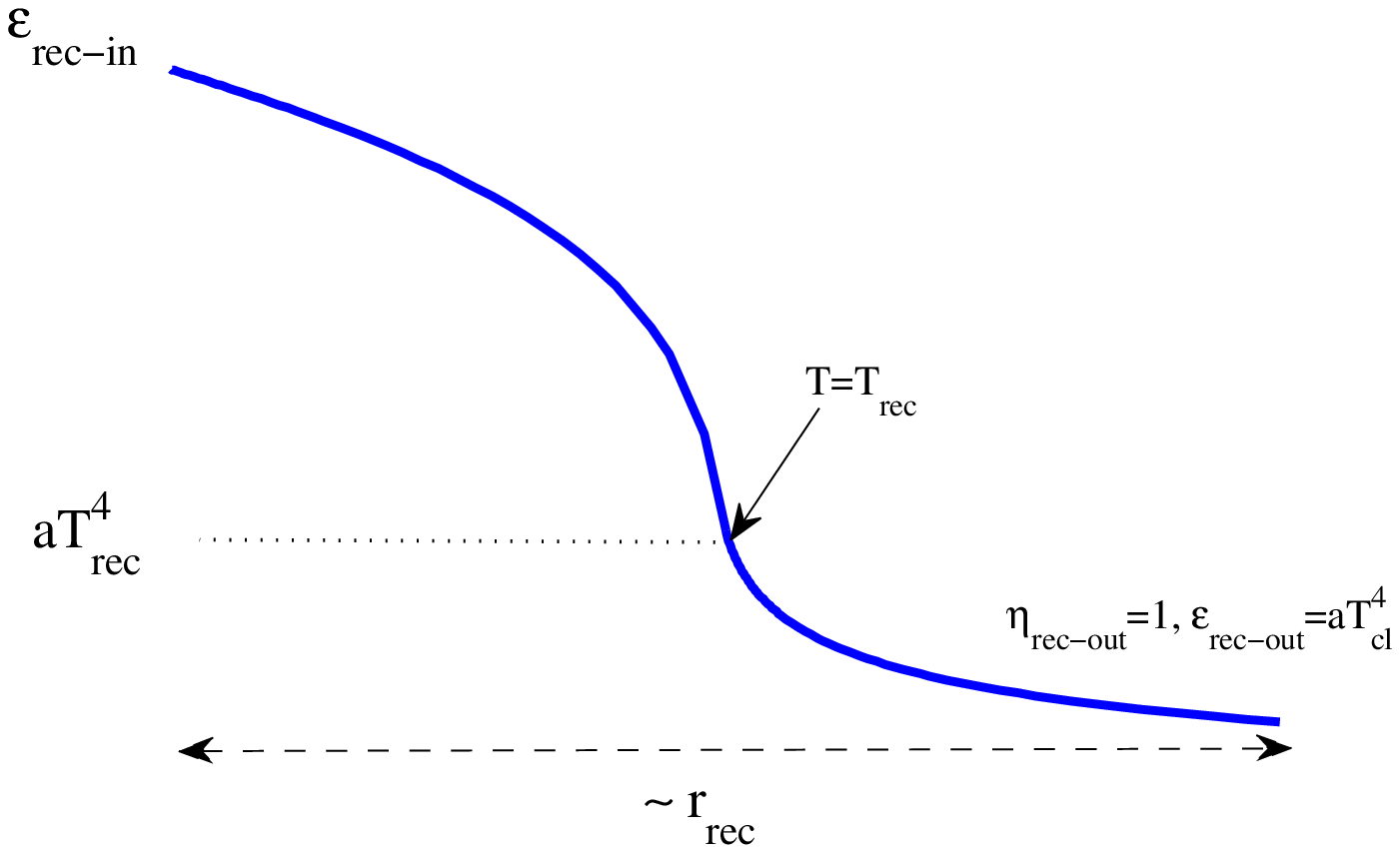}
\includegraphics[width=\columnwidth]{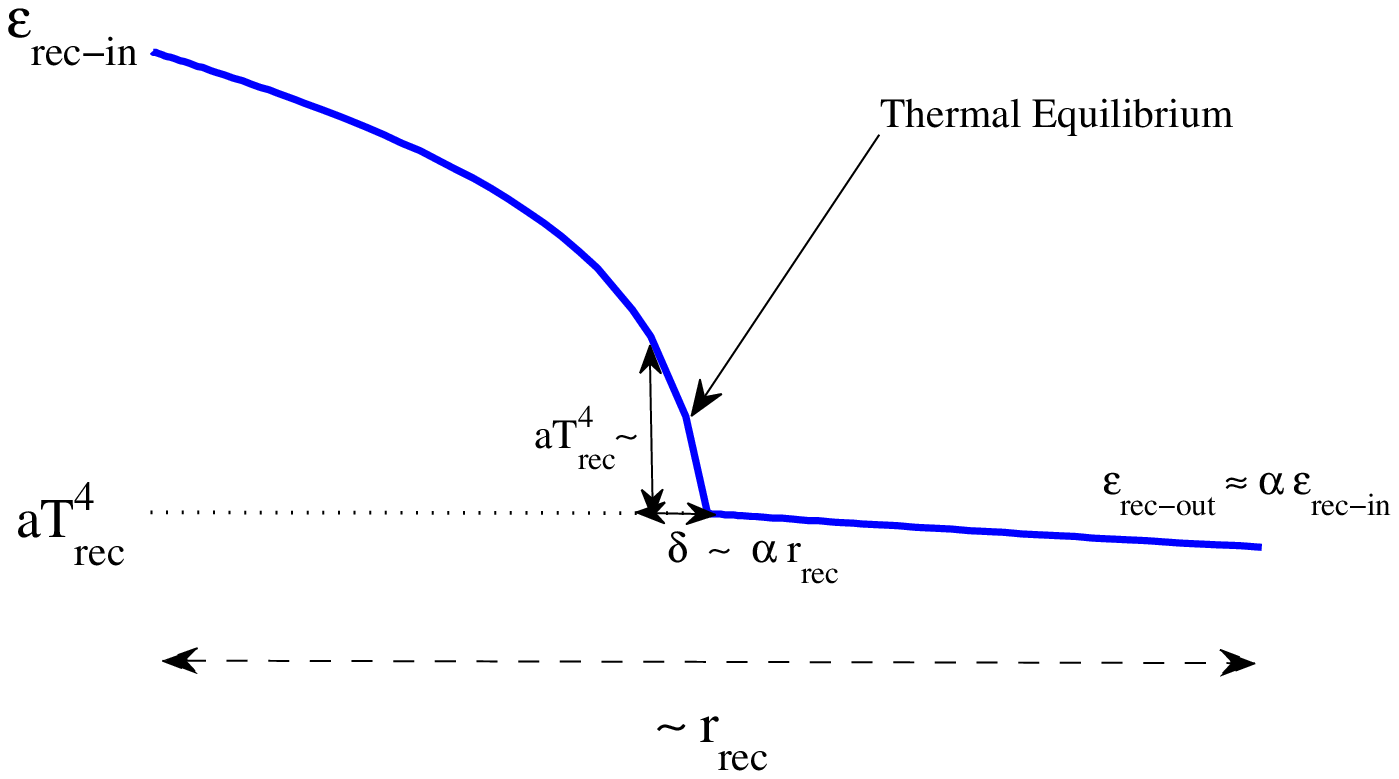}%
\caption{Schematic description of the energy density profile (logarithmic scale) inside the recombination shell 
when $m_{\rec}<\hat{m}$ for the two models- $x_{\ion} \propto T^{11}$ (\textbf{left}) according to~\S\ref{sec:mrec} and
$x_{\ion}$ is a step function (\textbf{right}) according to appendix~\ref{app:const}.
The internal part is in thermal equilibrium and its temperature $>T_{\rec}$.
The luminosity is dictated by an inner shell.
At the point where $T=T_{\rec}$ there is 
a sharp drop in the energy density because of the sharp decrease in the opacity at the external region.
In both cases the point where $T=T_{\rec}$ is in thermal equilibrium.
\textbf{In the left panel}: The temperature starts to flatten from the point where $T=T_{\rec}$ outward. The outer boundary is the outermost point which
is in thermal equilibrium.
\textbf{In the right panel}: The region infront of the point with $T=T_{\rec}$ is not necessarily in thermal 
equilibrium, see \ref{sec:temp_const} in appendix~\ref{app:const}.
Behind the point with $T=T_{\rec}$ on a scale of $\delta \sim \alpha r_{\rec}$ 
the energy density is approximately $a T_{\rec}^4$.}
\label{fig:mrec_out} 
\end{figure*}

Since $x_{\ion}$ is a continuous function of the temperature, the evolution 
is continuous (in contrast with the step function model discussed in the next subsection). 
As $m_{\rec}(t_{\rec})<\hat{m}(t_{\rec})$ the bolometric luminosity is not affected by recombination when it starts.
Only later, when the recombination wave arrives at the luminosity shell (so $m_{\rec}=\hat{m}=m_{\cl}$) the recombination wave determines the luminosity. We denote that time as $t_L$.

The recombination shell is found by solving equations \eqref{eq:general} for $m_{\rec}$ and 
$T_{\cl}$.
The expression for $\eta_{\rec-\out}$ is approximated with equation \eqref{eq:eta_r} using $m=m_{\rec}$,
$$
\eta_{\rec-\out}=
\left(\frac{T_{\cl}}{T_{\rec}} \right)^{\frac{7}{2}}
\left(\frac{\rho_{\rec}}{\rho_{\cl}(t_{\rec})} \right)^{-3} \times $$
$$
\left(\frac{r_{\rec}}{r_{\cl}(t_{\rec})} \right)^{-2} 
x_{\ion}(T_{\cl})^{-3} .
$$
We find
\begin{flalign*}
& m_{\rec}(t)=m_{\cl} (t) =&
\end{flalign*}
\begin{equation}
 \left\{
  \begin{array}{l l}
m_{\cl}(t_{rec})\left( \frac{t}{t_{\rec}}\right)
^{\frac{2(k-3)(-59ks+128k+177s-315)}{(k-2)(17k+87)}} &  
    \quad, t_{\rec} \leq t \leq t_L
     \vspace{0.5cm} \\
 m_{\rec}(t_L)
 \left( \frac{t}{t_{L}}\right)^
 {\frac{256(k-3)}{59ks+17k-177s+146}} &    
   \quad, t \geq t_L
  \end{array} \right.
  \label{eq:mrec}
\end{equation}
$$
\propto
 \left\{
  \begin{array}{l l}
     t^{2.26} &\quad, t_{\rec} \leq t \leq t_L 
     \vspace{0.5cm} \\
     t^{1.84} &\quad, t \geq t_L
  \end{array} \right. .
$$
Initially $m_{\rec}(t_{\rec})<\hat{m}(t_{\rec})$ and we can see that indeed the recombination wave is moving inward 
(in the Lagrangian sense) faster than the ionized luminosity shell (equation \ref{eq:lum_old}). It reaches the luminosity shell at
\begin{equation}
t_L=\Delta^{-\frac{(k-2)(17k+87)}{2(-59ks+111k+177s-402)(k-3)}}
t_{rec}
\approx \Delta^{-1.31} t_{rec} . 
\label{eq:tL}
\end{equation} 
At $t \geq t_L$ recombination takes place within the luminosity shell and determines its location, so $\hat{m}=m_{\rec}$.
The color temperature is given by
\begin{flalign*}
& T_{\cl} (t) =&
\end{flalign*}
\begin{equation}
 \left\{
  \begin{array}{l l}
T_{\rec} \left( \frac{t}{t_{\rec}}\right)
 ^{-\frac{2(k-3)(-6ks+11k+4s-14)}{(k-2)(17k+87)}} &  
    \quad, t_{\rec} \leq t \leq t_L
     \vspace{0.5cm} \\
 T_{\cl}(t_L) 
 \left( \frac{t}{t_{L}}\right)
 ^{-\frac{2(-7ks+11k+21s-26)}{59ks+17k-177s+146}} &    
   \quad, t \geq t_L
  \end{array} \right.
  \label{eq:Tcl}
\end{equation}
$$
\propto
 \left\{
  \begin{array}{l l}
    \propto t^{-0.17} &\quad, t_{\rec} \leq t \leq t_L 
     \vspace{0.5cm} \\
    \propto t^{-0.1} &\quad, t \geq t_L
  \end{array} \right. .
$$
We can see that once recombination starts the temperature evolves much more slowly
compared to its earlier evolution (equation \ref{eq:dawn_cl}), and it evolves even slower at $t>t_L$.

\begin{figure}[b!]
\includegraphics[width=\columnwidth]{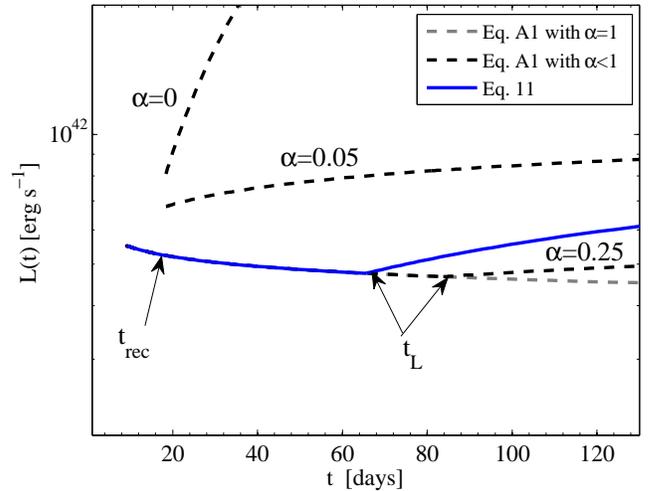}%
\caption{The bolometric luminosity for different approximations of the ionization fraction. The properties at $t_{\rec}$ are for a standard RSG progenitor ($M_*=15M_\odot$, $R_*=500R_\odot$)  and explosion energy of $10^{51}$ erg. After $t_{\rec}$ the hydrodynamic profile is approximated by $k=0.6$ and $s=0.95$. The blue solid line is for ionization model  
$x_{\ion} \propto T^{11}$. The black dashed lines are solutions for the step function models and 
the grey dashed line is for $\alpha=1$ where recombination has no effect.\\} 
\label{fig:lum_const_acc}
\end{figure}

\begin{figure}[t!]
\includegraphics[width=\columnwidth]{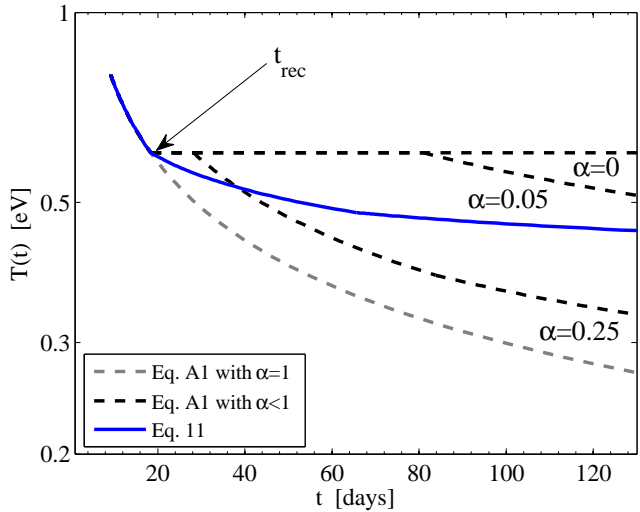}%
\caption{The observed temperature for different models. The initial conditions, hydrodynamic approximation and notations are the same  as in 
figure~\ref{fig:lum_const_acc}.
For the step function model, as the value of $\alpha$ is smaller
the observed temperature remains $T_{\rec}$ for longer times.\\} 
\label{fig:temp_const_acc}
\end{figure}

Finally, the bolometric luminosity, defined by equation \eqref{eq:num_lum} is 
\begin{flalign*}
& L (t) =&
\end{flalign*}
\begin{equation}
 \left\{
  \begin{array}{l l}
L(t_{\rec}) \left( \frac{t}{t_{\rec}} \right)^{-\frac{2(1-s)(k-3)}{k-2}} &  
    \quad, t_{\rec} \leq t \leq t_L
     \vspace{0.5cm} \\
 L(t_L)  \left( \frac{t}{t_{L}}\right)
 ^{\frac{2(69ks-17k-207s-18)}{59ks+17k-177s+146}} &    
   \quad, t \geq t_L
  \end{array} \right. 
  \label{eq:lum_mrec}
\end{equation}
$$
\propto
 \left\{
  \begin{array}{l l}
    \propto t^{-0.07} &\quad, t_{\rec} \leq t \leq t_L 
     \vspace{0.5cm} \\
    \propto t^{0.37} &\quad, t \geq t_L
  \end{array} \right. .
$$ 
At early times, when $t<t_L$ and $\hat{m}>m_{\rec}$, the luminosity does not depend on the dynamics of
the recombination shell. 
The increase in the luminosity for $t>t_L$ derived above could be expected as recombination
exposes more internal shells.
Examples for the above results are shown in figures~\ref{fig:lum_const_acc}
and ~\ref{fig:temp_const_acc},
where they are compared to the results regarding the step function model.

\subsection{Ionization Fraction as a Step Function} \label{sec:const}

Here we summarize the main results obtained when the ionization fraction is approximated by a step function. The detailed analysis is presented in appendix \ref{app:const}. As we argue below, this approximation is not physical. However, it emphasizes the importance of modeling the ionization fraction as a function which depends on the temperature when 
$T<T_{\rec}$ and it underlines some of the ideas behind the way we analyze the recombination shell.
In addition it can be compared with previous models which approximate the ionization fraction as a step function \cite[]{popov93,grass_nad76}.

The strength of the discontinuity in the ionization fraction
is given by the parameter $\alpha$ in equation \eqref{eq:alpha_const}. Just before recombination started the
color shell has the properties- $T_{\bb}(m_{\cl}(t_{\rec}),t_{\rec})=T_{\rec}$,
$\eta (m_{\cl}(t_{\rec}),t_{\rec})=1$ and it is also the recombination shell.
Because of the immediate change and the discontinuity in the opacity, at the onset of recombination $m_\rec$  is discontinuous.
In addition, because of the immediate drop in the free-free emission rate,
the mass of the color shell is also  discontinuous.
We find that from $t=t_{\rec}$ there is a transition time in which the recombination shell is 
also the color shell, but $\eta_{\rec-\out}>1$. 
Once $\eta_{\rec-\out}=1$, the color shell moves outward from the recombination shell.
Hence, equations \eqref{eq:general} are not relevant 
to the step function model and a more subtle analysis of the recombination shell profile is needed.   
  
There is a diversity of types of evolution which depend on the value of $\alpha$ with
respect to $\Delta$.
For example, 
we find a critical value $\alpha_L$ such that for $\alpha<\alpha_L$ there is a discontinuity in the luminosity 
at the onset of recombination. That is, when recombination starts it propagates on a very short time scale (shorter than the dynamical time) through the luminosity shell, exposing to the observer new regions. If $\alpha=0$, this discontinuity is followed by a significant increase in the luminosity.
There is also a range of values of $\alpha$ which results in a short period
of time where $m_{\cl}<m_{\rec}<\hat{m}$ so the propagation of the recombination shell does not affect
the luminosity nor the observed temperature.

For $\alpha \ll 1$ there is a sharp increase in the luminosity when recombination starts, see figure \ref{fig:lum_const_acc}.
This does not agree with observations of SNe.
On the other hand, for $\alpha$ that is not much smaller than 1, the temperature drops below $T_{\rec}$ short time after recombination starts,
see figure \ref{fig:temp_const_acc}.
The ionization fraction remains constant (since $\alpha$ is constant)
and therefore the physical conditions are not consistent, see figure \ref{fig:ross}. 


						
							

\section{Applying the Model For a Typical RSG Progenitor} \label{sec:application}

\subsection{General Properties of the Evolution} \label{sec:general}

We apply the analytical model introduced above to a general explosion.
We use the parameters $k=6$ and $s=0.95$ for the profiles of the envelope,
(see equations \ref{eq:vprof} and \ref{eq:eprof}).
The results are written as a function of the properties of the pre-SN and the
value of the recombination temperature.
According to the early evolution, see section \ref{sec:NS}, we find for the onset of recombination-
\begin{equation} \label{eq:trec_RSG} 
 t_{\rec} \approx 18 \text{ days }  M_{15}^{-0.78} R_{500}^{0.98} E_{51}^{0.51} T_{\rec,0.6}^{-2.43} ,
\end{equation} 
\begin{equation} \label{eq:delt_RSG}
 \Delta \equiv \delt \approx 0.38 \text{ } M_{15}^{0.05} R_{500}^{-0.18} E_{51}^{0.19} T_{\rec,0.6}^{1.11},
\end{equation}
where,
$M_{15}$ is the initial mass of the progenitor in $15 M_{\odot}$ units, $R_{500}$ 
is its initial radius in $500 R_{\odot}$ units, $E_{51}$ is the explosion energy in $10^{51} \text{ erg}$ units 
and $T_{\rec,0.6}$ is the recombination temperature in $0.6$ eV units.
We present here only results for the continuous ionization model (with $x_{\ion} \propto T^{11}$). Results for the step function model are presented in appendix \ref{app:const}.  

Using the results in~\S\ref{sec:mrec} we find 
\begin{equation}
t_L \approx 65 \text{ days }  M_{15}^{-0.85} R_{500}^{1.23} E_{51}^{0.26} T_{\rec,0.6}^{-3.9} ~.
 \label{eq:tL_RSG} 
\end{equation}  
The bolometric luminosity is affected by recombination only after $t_L$ and is given by 
\begin{flalign*}
& L(t) = &
\end{flalign*}
\begin{equation} \label{eq:lum_RSG}
   \left\{
  \begin{array}{l r}
   6.5 \times 10^{41} \text{ erg s}^{-1}   M_{15}^{-0.94} R_{500} E_{51}^{0.98} t_{\text{days}}^{-0.07}
    &  t_{\rec}<t<t_L\\
     \\10^{41} \text{ erg s}^{-1}   M_{15}^{-0.57} R_{500}^{0.45} E_{51}^{0.86} T_{\rec,0.6}^{1.73}  t_{\text{days}}^{0.37}
    &    t>t_L\\
  \end{array} \right. .
\end{equation}
On the other hand, the observed temperature changes its evolution with time from $t_{\rec}$, 
\begin{flalign*}
& T_{\cl}(t) =&
\end{flalign*}
\begin{equation} \label{eq:temp_RSG}
    \left\{
  \begin{array}{l r}
   0.98 \text{ eV}  M_{15}^{-0.13} R_{500}^{0.17} E_{51}^{0.09} T_{\rec,0.6}^{0.58}  t_{\text{days}}^{-0.17}
       &  t_{\rec}<t<t_L\\
   \\0.72 \text{ eV}   M_{15}^{-0.07} R_{500}^{0.08} E_{51}^{0.07} T_{\rec,0.6}^{0.87} t_{\text{days}}^{-0.1}
    &  t>t_L\\
  \end{array} \right. .
\end{equation}
The observed temperature is $T_{\rec}\approx 0.6$ eV at $t_{\rec}$ by definition, 
then it drops to $\Delta^{0.22}T_{\rec}\approx$ $0.5$ eV at $t_L$.
For $t>t_L$ it is almost constant with time.

\begin{figure}[t]
\includegraphics[width=\columnwidth]{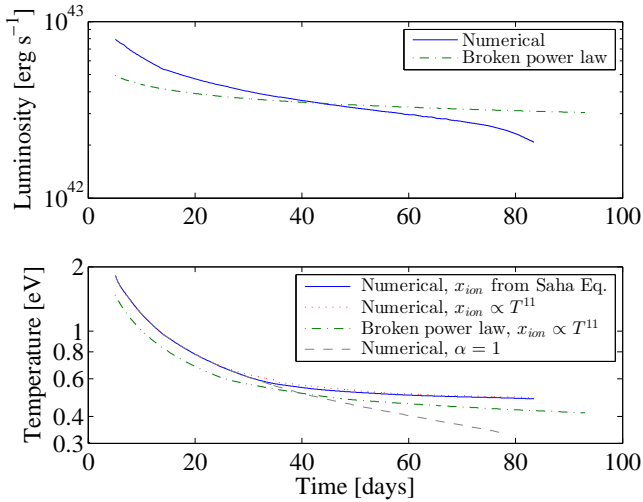}%
\caption{The bolometric luminosity and observed temperature for the two examples of RSG explosion described in \S\ref{sec:example}.
For the envelope taken from numerical simulation, there is no significant difference between the temperature whether
we parametrize ionization fraction with the analytical approximation, red dotted line,
or with more accurate numerical function according to Saha equation, blue solid line.
according to Saha equation.
The green dash-dotted line is for the envelope we described analytically using broken power law profiles.
The ionization fraction in that case is according to equation \eqref{eq:alpha} with  and 
$T_{\rec}=0.6$eV.  
In both examples recombination starts after day $\sim$30 and it does not affect the luminosity.
The grey dashed line described the solution for the numerical envelope with $\alpha=1$, where recombination has no effect
at all.\\} 
\label{fig:temp_lum_num}
\end{figure}

\subsection{Numerical and Analytical Examples}   \label{sec:example}

We use our  model to draw the light  curves for an explosion of  a typical RSG
progenitor.  We  present  two  models.  In   the  first  we  use  a  numerical
hydrodynamical profile of  a RSG explosion at the beginning  of the homologous
phase.  We  then  solve  the  radiation transfer  through  the  expanding  gas
semi-analytically based  on our recombination model. Second, we  use an
approximated  broken  power-law  for  the  hydrodynamics  and  solve  for  the
resulting               light                curve               analytically.

\begin{figure}[t]
\includegraphics[width=\columnwidth]{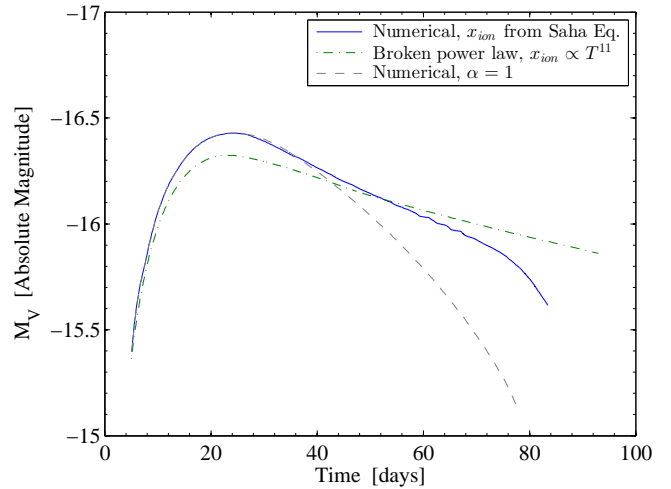}%
\caption{Optical $V$-band light curves ($5.44 \times 10^{14}$ Hz) for the two examples of RSG explosion described in \S\ref{sec:example}. 
The blue solid line is for the envelope taken from numerical simulation. The ionization fraction for that example is
according to Saha equation.
The green dash-dotted line is for the envelope we described analytically using broken power law profiles.
The ionization fraction in that case is according to equation \eqref{eq:alpha} with $T_{\rec}=0.6$eV.  
Recombination starts after day $\sim$30.
The grey dashed line described the solution for the numerical envelope with $\alpha=1$, where recombination has no effect.\\} 
\label{fig:V_const_acc}
\end{figure}

\begin{figure}[t]
\includegraphics[width=\columnwidth]{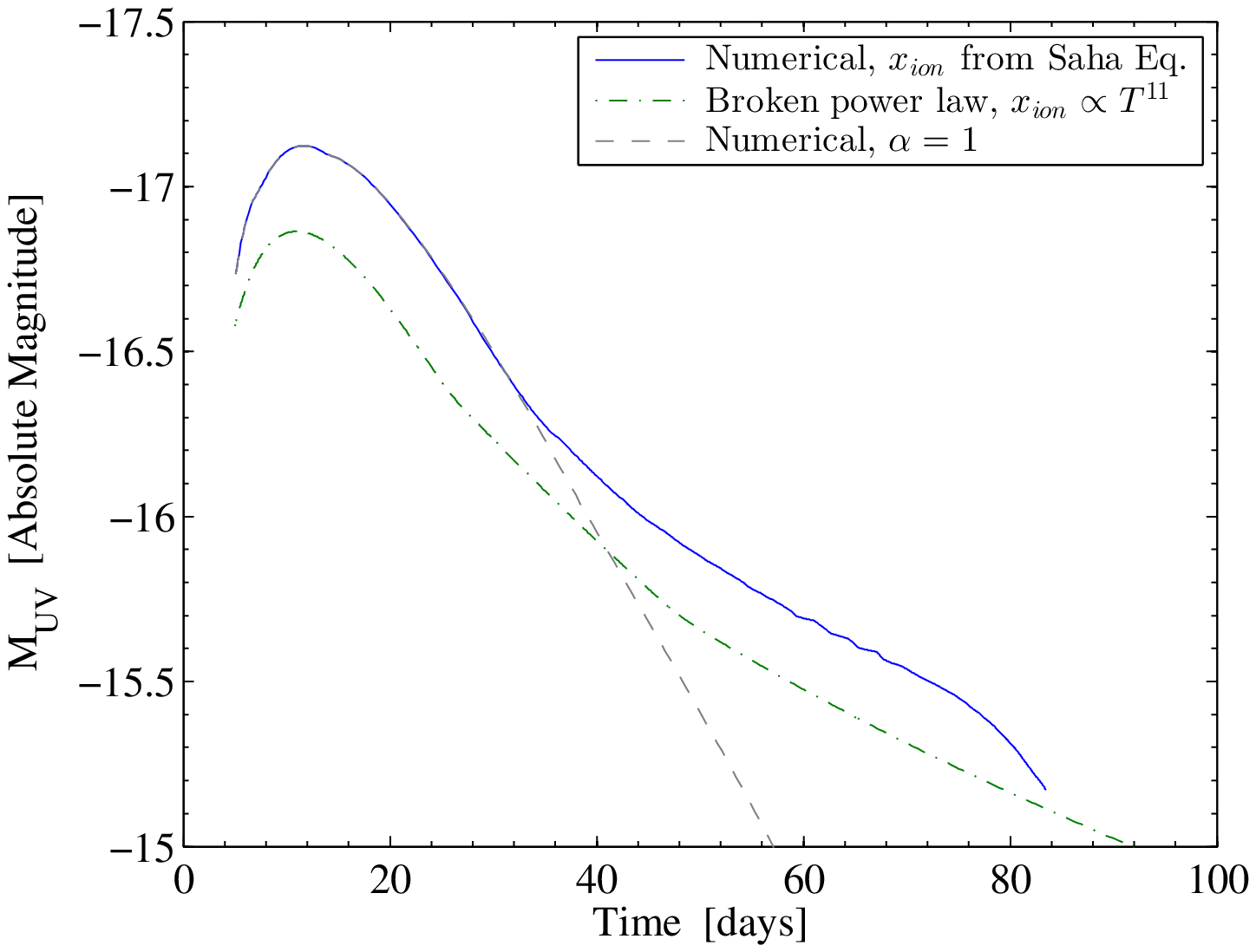}%
\caption{$u$-band light curves ($8.33 \times 10^{14}$ Hz) for the two examples of RSG explosion described in \S\ref{sec:example}.
The lines have the same meaning as in figure~\ref{fig:V_const_acc}.\\} 
\label{fig:UV_const_acc}
\end{figure}

\begin{figure}[t]
\includegraphics[width=\columnwidth]{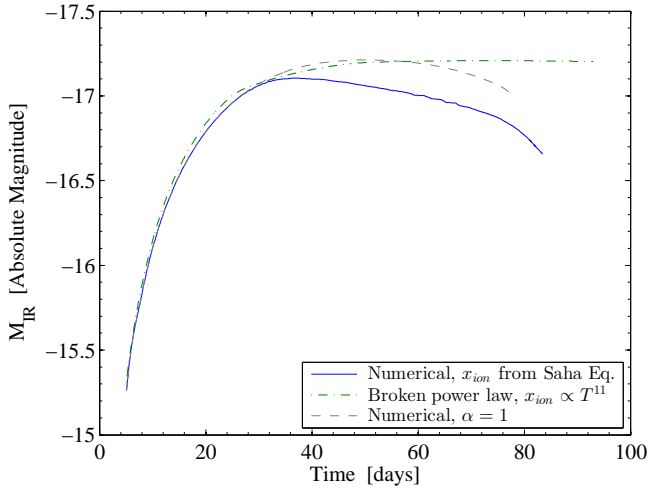}%
\caption{$I$-band light curves ($3.33 \times 10^{14}$ Hz) for the two examples of RSG explosion described in \S\ref{sec:example}.
The lines have the same meaning as in figure~\ref{fig:V_const_acc}.\\ } 
\label{fig:IR_const_acc}
\end{figure}

The numerical hydrodynamic profile of the  homologous envelope is taken from a
simulation of  a $M=13M_{\odot}$, $R=840 R_{\odot}$  progenitor explosion with
$E=1.2\times 10^{51}$ erg~\cite[]{Dessart2013}.  We use a snapshot
of the  hydrodynamic profile before  recombination starts at some  time during
the homologous expansion. Then we construct  the properties of the envelope at
later  times using  the fact  that  it continues  to expands  freely and  cool
adiabatically behind  the luminosity shell. The  light curve is then  found by
following the dynamics of the  luminosity shell, color shell and recombination
shell  during the  homologous expansion.  This  evolution is  governed by  the
general set  of equations  \eqref{eq:general}. These  equations can  be solved
numerically for  any $x_{\ion}(\rho,T)$,  $\rho(m)$, $E_{\ad}(m)$  and $v(m)$.
The  explicit   equations  we  solve   numerically  are  based   on  equations
\eqref{eq:general} with some correction of  pre-factors which were omitted and
definitions  which were  simplified in  the  analytic model  described in  the
previous  sections.  The   full  set  of  equations  are   given  in  appendix
\ref{App_num}.   We  solve   equations   \eqref{eq:num_equations}  until   the
luminosity shell passes  $8M_{\odot}$. From that point the  density profile is
roughly constant and our  model is not adequate. We consider  two different ionization
models  - (a)  $x_{\ion}(\rho,T)$  according to  Saha  equation (b)  $x_{\ion}
\equiv 1$ (i.e.,  no recombination). In figure  \ref{fig:temp_lum_num} we show
the  bolometric luminosity  and observed  temperature as  a function  of time.
These are  similar to those  observed in some type  IIp SNe. In  this specific
example the  color shell does not  reach the luminosity shell  before most the
trapped radiation escapes around day  90, implying that recombination does not
affect the  bolometric luminosity at all.  This is expected based  on equation
\ref{eq:tL_RSG}, which estimates  $t_L \approx 152$ day.  Thus, the bolometric
luminosity decreases very slowly with time. Recombination starts at $\sim$ day
30, and  from that  time the  temperature changes slowly  in time.  In figures
\ref{fig:V_const_acc},  \ref{fig:UV_const_acc}  and \ref{fig:IR_const_acc}  we
show    the     resulting    UV,     optical    and    IR     light    curves.

The second  model we present  is fully  analytic. The 
ionization  fraction  is approximated by a  broken  power-law, $x_{\ion} \propto T^{11}$ at $t<T_{\rec}=0.6$eV  (equation \ref{eq:alpha}). We  parametrize the envelope  of the progenitor as  a broken
power-law\footnote{ The expressions for  luminosity and temperature we provide
in \S\ref{sec:mrec} does not cover  the period in which  the luminosity
shell and color shell propagates  inward through different power-law profiles.
Nevertheless  the  relevant  expressions  can  be  readily  derived.}  in  the
following way. The  outer part was parametrize with power  law profiles, using
$k=12$ and $s=0.9$. We take this profile according to the profiles provided by
NS10  for  a RSG  with  $M=13M_{\odot}$,  $R=840 R_{\odot}$  and  $E=1.2\times
10^{51}$  erg.  This  outer  part  lies   from  the  surface,  inward,  up  to
$1M_{\odot}$.  From that  point  inward the  density  profile is  parametrized
continuously with $k=6$  and $s=0.95$. The reason for  this parametrization is
that  in the  numerical envelope  we use  for the  first example  we see  that
$s=0.95$ is a good approximation for  any $1M_{\odot}<m$ while $k=6$ is a good
approximation in the region $1M_{\odot}<m<3M_{\odot}$. At larger masses k drops
continuously (the density profiles becomes shallower).  For that  reason the  analytic solution  provides only  a rough
approximation to  a realistic RSG explosion.  Nevertheless, the characteristic
behavior of the  bolometric luminosity and observed temperature  is similar to
that of the numerical hydrodynamic profile, see figure \ref{fig:temp_lum_num}.

The optical and the IR light curves in this example show a slow evolution in time from the onset of recombination at $\sim$ day
30. This is in agreement with the plateau feature (up to 0.5 in magnitude) seen in observations. The UV light however, which is much more sensitive to small changes in the temperature, drops much faster. Again with agreement with observations. In this example the bolometric luminosity is not affected by recombination. But, in other cases (more compact and/or massive progenitors) the recombination wave is expected to reach the luminosity shell during the plateau phase. A very slow evolution of the optical and IR bands is expected also in that case, with the main difference that the optical light may rise slowly. On the other hand, the slower evolution of the temperature will have a strong effect on the UV light curve which is expected to fall much more slowly, or even rise very slowly, once recombination affect the luminosity. This may explain the late UV plateau observed in some SNe \cite[]{Bayless13}.


\subsection{Plateau Luminosity and Duration} \label{sec:scaling}

Over the years, much effort has been made in order to connect
the observable data to the progenitor and the explosion properties.
In type IIP SNe, there have been attempts to relate three observed properties of the SN light curve - the plateau duration, plateau luminosity
and photosphere velocity - to three physical properties -  the progenitor  mass and radius and the explosion energy.
Such relations were derived analytically by \citet{arnett80} who ignores recombination and by~\citet{popov93}
who uses a step function parameterization with $\alpha=0$. Numerical analysis by~\citet{litvinova85} gives results similar to those
of~\citet{popov93}.

The plateau duration, $t_{\text{SN}}$, can be approximated by the time in which the energy deposited by the shock is released from the \textit{entire} envelope. 
We examine the energy release by the ``last" shell, constituting the bulk of the star. It has a mass $\sim M_{\text{ej}}$, radius $r=v_{\sn}t_{\sn}$
and density $\sim M_{\text{ej}}/r^3$, where $M_{\text{ej}} \sim M$ is the ejected mass and $v_{\sn} \approx \sqrt{2E/M_{\text{ej}}}$.
If recombination does not affect the luminosity (i.e., $t_L>t_{\text{SN}}$), the relations derived by \citet{arnett80} are valid:
\begin{equation}
	\begin{array}{lll}
		t_{\text{SN}} &\propto& \frac{M^{0.75} \kappa^{1/2}} {E^{0.25}}  \\
		&&\\
		L_{\text{SN}} &\propto& \frac{E R} {M \kappa}  
	\end{array}
	~~~~t_{\text{SN}}<t_L,
\label{eq:lsn_Arnett}
\end{equation}
where $L_{\sn}$ is the typical luminosity of the plateau

When recombination affects the luminosity, the plateau luminosity and duration can be determined by the following relation
$$
E_0 \left( \frac{r}{R_{\star}} \right)^{-1} t_{\sn}^{-1} \approx L_{\sn} \approx
\frac{a T_{\cl}^4 r^3} {x_{\ion}(T_{\cl}) \kappa_T \rho r^2} ,
$$
where $E_0 \approx E/2$ is the initial internal energy. We use the ionization model according to equation \eqref{eq:alpha} 
so $T_{\cl}$ and $x_{\ion}(T_{\cl})$ should be found 
by another equation regarding the thermal coupling coefficient of that color shell (equation \ref{eq:num_eta}).
Using the above definitions, we find
\begin{equation}
	\begin{array}{lll}
		t_{\text{SN}} &\propto& \frac{M^{0.45} R^{0.23}  
		\kappa^{0.76}} {E^{0.15} T_{\rec}^{0.73}}\\
		&&\\  
		L_{\text{SN}} &\propto& \frac{ E^{0.81} R^{0.54}T_{\rec}^{1.46} 
		} {M^{0.4} \kappa^{0.35}}  
	\end{array}
	~~~~t_{\text{SN}}>t_L.
\label{eq:ltsn}
\end{equation}
The general behavior of the scalings we find is rather similar to previous analytical and numerical results (for comparison with scalings from other works see section \ref{sec:comparison}). Therefore, if $E,R$ and $M$ are independent, the variation in $t_{\text{SN}}$ is expected to be larger than the observed, almost constant, plateau duration of 100 days \cite[]{arcavi12}. This implies that there is probably a dependence between these explosion properties \citep{Poznanski13}.

\section{Comparison to Previous Works} \label{sec:comparison}

\subsection{Analytical Works}

A couple of works have delt with the problem of recombination in SNe envelopes.
These works examined a recombination wave which runs inward through the envelope 
and dictates the energy emission\footnote{
Some works use the term ``wave of cooling and recombination'' (WCR) or ``cooling and recombination wave'' (CRW).}.  
In these works, recombination divides the envelope into two regions- 
hot and opaque behind the recombination front, cold and transparent in front of it.
The transparent region is characterized with a temperature $T_2$ and $\tau(\rho,T_2)=1$. The luminosity is
$L=4\pi R_{\ph}^2 \sigma T_2^4$, where $\sigma$ is the Stephan-Boltzmann constant, and $R_{\ph}$ is the radius of
the recombination front which is also the photosphere radius (defined as the point where $\tau=1$).
Thus, the main difference between our work and previous ones is that we examine the coupling between radiation and matter while previous works assume full coupling.
In our model  we find that $\tau_{\rec-\out}(t)>1$ for a typical RSG explosion,
so the photosphere is located further out
from the recombination shell. Moreover, since the recombination shell is also the outermost point in thermal equilibrium, 
the relation $L=4\pi R_{\ph}^2 \sigma T_2^4$ is not valid.   
In addition, we show that recombination does not affect the luminosity at the onset of recombination.
This also comes from the account for coupling between radiation and matter, or more explicitly, the
separation between the luminosity shell and the color shell at early times.
In the rest of this subsection, we discuss more specifically two fundamental previous works by \cite{grass_nad76}
and \cite{popov93}
which addressed the problem of recombination in SN envelope.

\citet{grass_nad76} examined a recombination wave propagating in a SN envelope.
The homologous expanding envelope has a density profile $\rho \propto r^{-k} t^{k-3}$, as in our model.
They gave a significant role to the energy released from the recombination itself in
determining the temperature at the inner region of the recombination wave.
Here we shortly summarize their results, focusing on the recombination shell and its boundary conditions.
The energy in the internal boundary of the recombination shell is given by $\chi \rho_{\rec} /m_{\mu}$, where $\chi$
is the recombination potential and $m_{\mu}$ is the average mass per ion. The diffusion time in the external boundary
is $\sim r_{\rec}/c$, since they assume $\tau_{\rec-\out}=1$. Equating the luminosity form the external boundary to
the release of the energy in the internal boundary during the dynamical time of the shell ($\sim t$)
gives $$\rho/t \propto T_2^4/r_{\rec}.$$ 
Under the assumption that $T_2\approx$ constant, one finds that $r_{\rec} \propto t^{\frac{k-4}{k-1}}$.
Since in this analysis $L \propto r_{\rec}^2$, then for example $L \propto t^{0.8}$ for $k=6$. 
In contrast, we find that the energy from the recombination itself does not play any significant role in
determining the dynamics of the recombination shell. Let us consider late times, when $m_{\rec}=\hat{m}$.
The luminosity released from the internal boundary of the recombination shell, including the energy from recombination, is
$L=E_{\ad}(m_{\rec})/t+\chi \rho_{\rec} r_{\rec}^3 /m_{\mu}t$. Thus, the recombination energy is important
when 
$$\frac{\chi}{k_B T_{\bb,\rec-\in}} \beta_{\rec-\in} \gg 1,$$
where $\beta_{\rec-\in}$ is the ratio between gas pressure and radiation pressure in the internal boundary of the shell.
We find that this condition is not satisfied, 
$\beta_{\rec-\in} \sim \frac{k_B T_{\rec-\in} \rho_{\rec}}{a T_{\rec-\in}^4 m_{\mu}} \sim 10^{-2} 
\left( \frac{k_B T_{\rec-\in}}{\text{eV}} \right) ^{-3}$.
 It is only if $T_{\rec-\in} \sim T_{\rec}$ that this 
inequality may hold. However, this situation does not occur in our solution.
In addition, \citet{grass_nad76} found
typical values of $x_{\ion}=10^{-4}-10^{-3}$ which is lower than the typical values of $x_{\ion} \geq 10^{-2}$ found by us and by detailed numerical simulations (see below).       

\citet[]{popov93} considered a homogeneous density in the freely expanding envelope.
He used an analytic solution by \citet[]{arnett80} who ignores recombination ($\kappa=\kappa_T$).
This solution sets conditions before recombination starts that are not realistic in the actual case of a non-homogeneous envelope. The opacity in Popov's model is a step function, such as in equation
\eqref{eq:alpha_const}, with $\alpha=0$. He sets the observed temperature to be $T_2=T_{\rec}$ from the onset of recombination
so the recombination wave has a zero width\footnote{
Popov's model does not deal with color temperature and coupling between matter and radiation, so the onset of 
recombination is defined by the time when the temperature at the photosphere is $T_{\rec}$.}.  
We find that if $\alpha=0$, then at the onset of recombination there is a sharp increase in the luminosity 
(follows a discontinuity in the evolution). This phenomena is not seen in Popov's solution. 
However, we can not give a fair comparison between the evolution of the luminosity and observed temperature derived above
and those derived by Popov, since he treated a problem with homogeneous density which we do not consider here. Nevertheless, the relations that Popov finds for the plateau duration  and luminosity are not very different than our relations when recombination plays a role:
\begin{equation}  
	\begin{array}{lll}
		t_{\text{SN,Popov}} &\propto& \frac{M^{1/2} R^{1/6}  
		\kappa^{1/6}} {E^{1/6} T_{\rec}^{2/3}} \\
	&&\\  
		L_{\text{SN,Popov}} & \propto& \frac{ E^{5/6} R^{2/3} T_{\rec}^{4/3} 
		} {M^{1/2} \kappa^{1/3}}  
	\end{array} .
\label{eq:tsn_popov}
\end{equation}

\subsection{Numerical Works}
   
Our results regarding  the propagation of a recombination wave  in the ejected
envelope of a RSG  are compatible  with the picture that is found in
numerical  studies. Those  studies examine  the photometric  and spectroscopic
evolution of SNe, from  the initial pulse up to the  nebular phase. They find,
similarly to our model, that recombination takes place over a small mass scale
(much smaller than $m_{\rec}$). The  radiation and gas temperature are coupled
in the  ionized region and they  remain coupled while they  drop significantly
within the  recombination shell.  Once the temperature  falls to  the observed
temperature the gas  and the radiation decouple and the  gas continues to cool
while  the   radiation  temperature  is   constant  \cite[e.g.,][]{utrobin07}.
Quantitatively,  the main  result of  our model  that can  be compared  to the
published results of numerical simulations is the ionization level at the point
that    the   radiation    observed   temperature    is   determined    (i.e.,
$x_{\ion}(T_{\cl})$).  For typical  RSG parameters  our model  (using equation
\ref{eq:alpha}) predicts $x_{\ion}(T_{\cl}) \approx 0.05$ at $t=100$ day. This
value agrees  with numerical  models that  properly include  the radiation-gas
coupling  \cite[e.g.,][]{utrobin07,woosley&kasen09},   which  find   that  the
ionization fraction  at the  external boundary of  the recombination  shell is
$\approx  10^{-2}-10^{-1}$.  It  also  explains the  physical  origin  of  the
``opacity  floor'', which  is  invoked in  order to  fit  the observations  in
numerical   simulations   that   do    not   treat   the   coupling   properly
\citep[e.g.,][]{Young04,bersten11}. The typical  value for this floor  in hydrogen rich
material is  $0.01 {\rm~cm^2~gr^{-1}}$, which corresponds  to $x_{\ion}=0.03$.
Another related prediction  of our model is  that the observed temperature  falls from
$T_{\rec}=6,500-7,000$   K  at   the   time  that   recombination  starts   to
$5,000-6,000$  K by  the time  that it  ends. This  is similar  to the  values
obtained                in                numerical                simulations.

\section{Summary and Discussion} \label{sec:summary}

We study analytically the effect of recombination on the evolution of SNe light curves.
The process of recombination reduces the number of free electrons, reducing both the opacity and the radiation-gas coupling (via the ability of the gas to absorb and emit photons). The balance between the two, opacity and coupling, is what determines the radiation temperature and ionization level in the recombined gas. Once recombination starts 
a wave of recombination, which is followed by cooling, starts to propagate inward through the envelope. 
 
We develop  a model which describes  the properties of the  relevant shells of
the envelope-  the shells  which dominates the  bolometric luminosity  and the
observed  temperature.  The  concept  of   the  model  is  to  consider  three
characteristic  shells which  dominate the  light curves:  (1) The  luminosity
shell, which is the source of the  observed luminosity. This shell determines the
radiative flux  through all  the shells  external to it.  (2) The  color shell
which is  the outermost  shell in  which photons can  be thermalized  over the
diffusion time. This shell dictates the color of the observed temperature. (3)
The recombination shell which is the shell where recombination takes place. It is
the outermost  shell which is  fully ionized on its  inner boundary and  it is
highly recombined on  its outer boundary. We proved that this shell
coincides with
the color shell once recombination starts and at late times it also coincides
with the luminosity shell. The main  advantage of our model
over previous analytical studies of the  process is that it takes into account
the coupling between the  radiation and the gas, which is  found to be crucial
in finding the ionization level at the point where the observed temperature is
determined. It can, therefore, provide a reliable estimate of the evolution of
the observed temperature and of  the bolometric luminosity. Our model provides
new insight  into the recombination process  and to the relations  between the
observables     and    the     progenitor     and    explosion     properties.

We find that when recombination starts  it first affects only the temperature,
which becomes almost  constant with time. The bolometric luminosity,  at first, continues
to drop  very slowly without being  affected by recombination. Only  later the
recombination wave  reaches the  luminosity shell.  The temperature  from that
point evolves even more slowly, while the bolometric luminosity stops dropping
or even  rises slowly.  The effect of  this transition is  rather mild  in the
optical but it should cause significant flattening of the UV light. Whether
the recombination wave  reaches the  luminosity shell before  the end  of the  plateau depends  on the
progenitor  properties (e.g., it  is  more likely  to take  place  in less  extended
progenitors). We farther  find that the radiation-gas coupling  limit the drop
of the  ionization fraction in  the recombination  shell to 0.01-0.1.  This in
turn is the origin of the very slow drop in the observed temperature, which is
$T_{\rec}=6,500-7,000$ K  when recombination  starts and $5,000-6,000$  K when
the                                plateau                               ends.

Another interesting result of our analysis is that the observed plateau is not
a generic  property of the  propagation of  a recombination wave  in expanding
ionized gas. The generic property is the much slower evolution of the observed
temperature  once recombination  starts (due  to  the strong  feedback of  the
temperature drop on the coupling).  This fixes the observed temperature around
the  $R$  and  $I$  bands.  However, the  slow  evolution  of  the  bolometric
luminosity is  a result of  the typical hydrodynamical structure  of exploding
stellar envelopes. Different  structures can result in a much  faster decay or
rise of  the bolometric luminosity.  Slightly different stellar  structures is
probably the main source of the  various behaviors seen in  type IIp SNe.

The derivation of  a fully analytical  solution requires the following approximations- 
(1) power-law profile for the density and adiabatic cooling at late times; 
(2) neglecting the density dependence in the ionization fraction;
(3) power-law relation between the ionization fraction and the temperature;
(4) neglecting radioactive energy from $^{56}$Ni;
(5) assuming free-free as the most efficient process for photon production.
However, the basic ideas of our model- (1) separation of the envelope into successive shells; (2) finding the recombination shell
by the properties of its boundaries, can be used for more general assumptions than (1)-(6) above, as is described in 
appendix \ref{App_num}.
The simple equations introduced in \eqref{eq:num_equations} can be extended even further by changing the function
$\eta$ such that it includes bound-free process or by adding energy from radioactive decay.
Hence, our model serve as a simple tool to examine the effect of some basic mechanisms on
the evolution of SNe light curves.
 
 \acknowledgments
 This research was partially supported by an ERC starting grant, ISF and ISA grants, and an iCore center.
 We thank Roni Waldman and Eli Livne for providing us an example of a numerical stellar envelope.


												

\appendix

\section{Ionization Fraction as a Step Function} \label{app:const}

Consider the following parametrization
\begin{equation} \label{eq:alpha_const}
   x_{\ion} = \left\{
  \begin{array}{l l}
    1 & \quad   T>T_{\rec}\\
    \alpha  & \quad T<T_{\rec}\\
  \end{array} \right. ,
\end{equation}
where $0 \leq \alpha<1$.
As opposed to the description in \S\ref{sec:rec_shell}, now the opacity does not drop with the temperature for 
$T<T_{\rec}$. Therefore the recombination shell is not necessarily coupled to the color shell
and equations \eqref{eq:num_lum}-\eqref{eq:num_eta} are not valid.
The dynamics of the luminosity shell is still dictated by equation \eqref{eq:num_mhat}
and the color shell is simply defined by
\begin{equation}
\left \{
  \begin{array}{l l}
   \eta(m_{\cl},t)=1 & \quad   \text{, if } m_{\cl}<m_{\rec}\\
   m_{\cl}=m_{\rec} & \quad   \text{, otherwise}\\
  \end{array} \right. .
\label{eq:general_const_cl}
\end{equation}
In \S\ref{sec:mrec_const}, we find the dynamics of the recombination shell and
by doing so, we also find the evolution of the bolometric luminosity.
In \S\ref{sec:temp_const} we discuss the dynamics of the color shell and find the evolution 
of the observed temperature for the different scenarios.  
It is quite cumbersome to describe all the different scenarios of the evolution.
Nevertheless, the description below gives a complete picture and the main results and equations.
In order to illustrate the various scenarios 
we summarize the evolution in table~\ref{table1}.
In the following analysis 
we use the notation $t_{\rec}^{\pm}$ to emphasize the discontinuity at the onset of recombination.
It is evident from the following analysis that the dynamics of the recombination shell and the color shell are continuous in $\alpha$. 
Thus, for any choice of $\alpha$, a unique solution for the evolution is found.

\begin{figure}
\includegraphics[width=0.5\columnwidth]{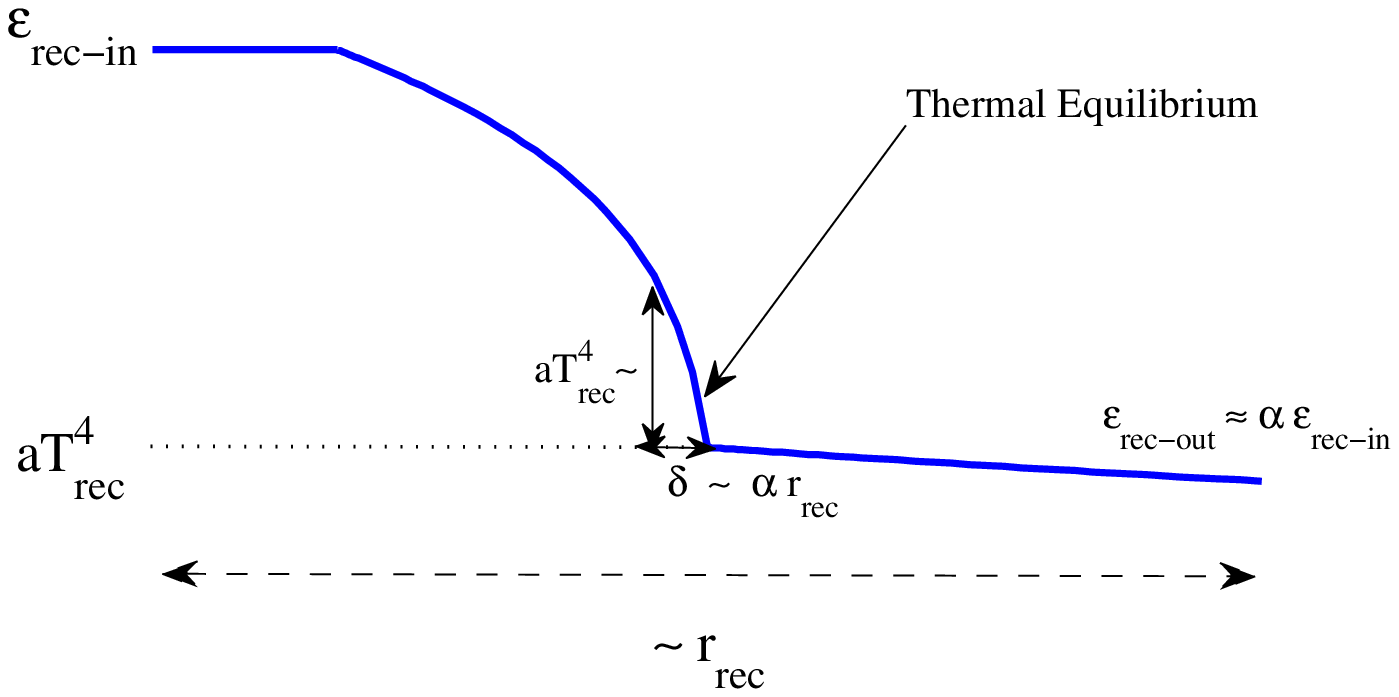}%
\includegraphics[width=0.5\columnwidth]{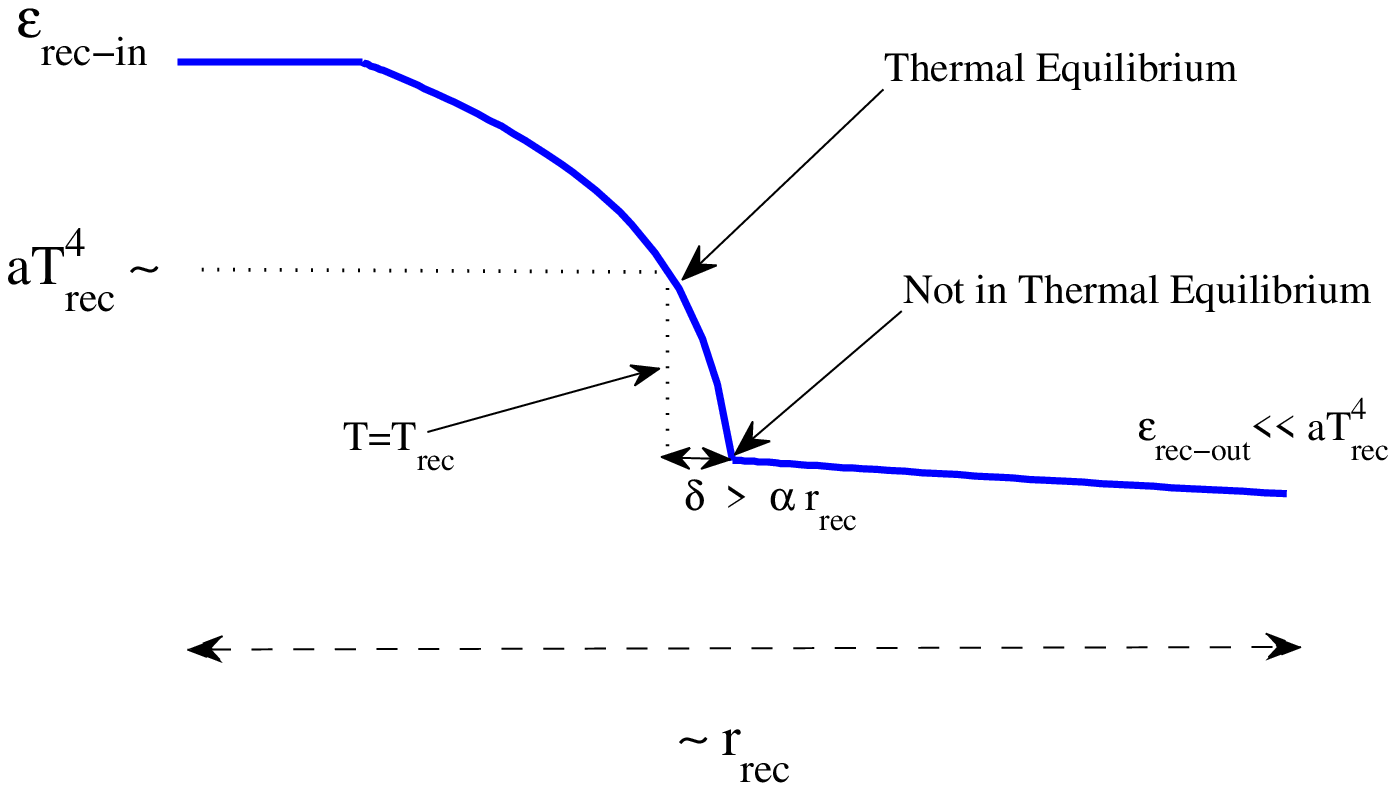}%

\caption{Schematic description of the energy density profile (logarithmic scale) inside the recombination shell 
when $m_{\rec}=\hat{m}$ for the step function model, shown for two different scenarios:
When The radiation behind $T=T_{\rec}$ is in thermal equilibrium (\textbf{left}) and
when it is not (\textbf{right}), according to~\S\ref{sec:mrec_const}.
\textbf{In the left panel}: Behind the point with $T=T_{\rec}$ in a subshell with a width of $\delta \sim \alpha r_{\rec}$ 
the energy density is approximately $a T_{\rec}^4$. 
The external boundary is not necessarily in thermal 
equilibrium, see \S\ref{sec:temp_const}. However, the energy density there is $\sim a T_{\rec}^4$. 
\textbf{In the right panel}: The radiation departs from thermal equilibrium in a subshell with a width of $\delta > \alpha r_{\rec}$
located behind the point where $T=T_{\rec}$. 
The energy density in this subshell is approximately $a T_{\rec}^4$. 
From the point in which the radiation departs from thermal equilibrium out to the point where $T=T_{\rec}$,
the energy density decreases significantly while the temperature is roughly constant $\sim T_{\rec}$.
From the point with $T=T_{\rec}$ outward the profiles of the energy density and the temperature are
approximately homogeneous.
In both cases, 
the luminosity in the discussed subshells is determined
by the luminosity in internal region.\\}  
\label{fig:mrec_in_const}
\end{figure}

\subsection{Dynamics of the Recombination Shell } \label{sec:mrec_const}
  
Consider the recombination shell at $t>t_{\rec}$.
In the internal boundary we have
$\eta_{\rec-\in}<1$ and $\epsilon_{\rec-\in}>a T_{\rec}^4$. 
As we go further out, but still within the shell, toward the point where $T=T_{\rec}$, the energy density
and the thermal coupling decrease significantly.
For $T \leq T_{\rec}$ all the hydrodynamic properties changes on a scale of
$r_{\rec}$, thus, from the point where $T=T_{\rec}$ up to the external boundary
they can be approximated as homogeneous.
Schematic profile inside the recombination shell is given in figures \ref{fig:mrec_out} and \ref{fig:mrec_in_const}.
We need to understand the conditions at a small subshell behind the point with $T=T_{\rec}$.
The width of this subshell is the scale on which there is no significant change in the characteristic of this subshell.
Let us take a distance $\delta$ inward from the point with $T=T_{\rec}$, and consider a subshell with width $\delta$.
The optical depth and the diffusion time of this subshell are given by
\begin{equation}
\tau(\delta)=\kappa_T \rho_{\rec} r_{\rec} \left( \frac{\delta}{r_{\rec}} \right) \quad , \quad
t_d(\delta)=\frac{\kappa_T \rho_{\rec} r_{\rec}^2}{c} \left( \frac{\delta}{r_{\rec}} \right)^2 .
\label{eq:tau_delta}
\end{equation}
According to equation \eqref{eq:eta_def}, the thermal coupling coefficient of this subshell is 
\begin{equation}
 \eta(\delta)=
\left(\frac{\epsilon(\delta)}{\epsilon_{\cl}(t_{\rec}^-)} \right)^{7/8}
\left(\frac{\rho_{\rec}}{\rho_{\cl}(t_{\rec}^-)} \right)^{-3}
\left(\frac{r_{\rec}}{r_{\cl}(t_{\rec}^-)} \right)^{-2}
\left( \frac{\delta}{r_{\rec}} \right)^{2} .
\label{eq:eta_delta}
\end{equation}
Note that we consider only subshells with $\delta \gtrsim \alpha r_{\rec}$.
The subshell with $\delta \sim \alpha r_{\rec}$ is the smallest subshell which can be considered
with homogeneous properties.
The luminosity in the external boundary of the recombination shell and in the subshell is dictated by internal regions.
The equation which governs the dynamic of the recombination shell is
\begin{equation}
\frac{\epsilon_{\rec-\out} c r_{\rec}}{\alpha \kappa_T \rho_{\rec}}=
\frac{E_{\ad}(\hat{m})}{t}=
\frac{\epsilon(\delta) r_{\rec}^3 \left( \frac{\delta}{r_{\rec}} \right) } 
{t_d(\delta) } ,
\label{eq:general_const_rec}
\end{equation}
where $\delta$ or $\epsilon_{\rec-\out}$ should be determined.
If the change in the opacity is not too large then the radiation 
is in thermal equilibrium at least up to the point where $T=T_{\rec}$. More precisely, 
there is a critical value, denoted with $\alpha_0$, such that for $\alpha>\alpha_0$ this is true 
at the onset of recombination.
Hence, we divide our analysis of the dynamics of the
recombination shell in to two different scenarios- 
when the radiation behind $T=T_{\rec}$ is in thermal equilibrium and when it is not.
We emphasize here that the condition of thermal equilibrium behind $T=T_{\rec}$ means
that the energy density at the point where $T=T_{\rec}$ is $aT_{\rec}^4$.  

Consider the first scenario,   
from the continuity of the energy density, we know
that the energy density at the external region is also roughly $a T_{\rec}^4$. Hence, 
$\epsilon_{\rec-\out} \approx a T_{\rec}^4$.
Using this to solve the left equality in equation \eqref{eq:general_const_rec} for $m_{\rec}(t)$ 
we find
\begin{equation}
m_{\rec}(t)=
 \left\{
  \begin{array}{l l}
   m_{\cl}(t_{\rec}^-) \alpha^{-\frac{k-3}{k+1}}
	\left( \frac{t}{t_{\rec}} \right)
	  ^{\frac{2(k-3)(-(k-3)s+3k-7)}{(k+1)(k-2)}} &  
    \quad, m_{\rec}<\hat{m}
     \vspace{0.5cm} \\
  \hat{m}(t_{\rec}^-)
     \left( \frac{\alpha}{\alpha_L} \right)^{-\frac{k-3}{(k-3)s+k+2}}
    \left( \frac{t}{t_{\rec}} \right)^{\frac{6k-18}{(k-3)s+k+2}} &    
   \quad, m_{\rec}=\hat{m}
  \end{array} \right.
\propto
 \left\{
  \begin{array}{l l}
    \propto t^{1.75} &\quad, m_{\rec}<\hat{m} 
     \vspace{0.5cm} \\
    \propto t^{1.66} &\quad, m_{\rec}=\hat{m}
  \end{array} \right. ,
\label{eq:mrec_const}
\end{equation}
where
\begin{equation}
\alpha_L= \Delta^{ \frac{k+1}{k-3}} 
\approx \Delta^{ 2.33 } .
\label{eq:alphaL}
\end{equation} 
As is expected, at $t_{\rec}$ there is a discontinuity in the recombination shell mass and afterwards
recombination continues to reach deeper shells.
If the change in the opacity is large, that is $\alpha$ is lower than the critical value $\alpha_L$,
then the recombination shell can affect the luminosity from the onset of recombination.
This scenario is defined by the condition $m_{\rec}(t_{\rec}^+)>\hat{m}(t_{\rec}^-)$. 
This condition implies that the recombination
shell is also the luminosity shell from the onset of recombination. 
If $\alpha>\alpha_L$, the recombination shell moves rapidly inward, so finally at 
some time $t=t_L$ it reaches to the luminosity shell.
The value of $t_L$ is defined by $m_{\rec}(t_L)=\hat{m}(t_L)$, where $m_{\rec}$ 
is given by the first expression in equation \eqref{eq:mrec_const},
\begin{equation}
t_L=\left(\frac{\alpha} {\alpha_L} \right) ^{-\frac{k-2}{2((k-3)s-2k+8)}}
t_{\rec} 
\approx
\left(\frac{\alpha} {\alpha_L} \right) ^{1.74}
t_{\rec} .
\label{eq:tL_const}
\end{equation}     
The expression for the bolometric luminosity is
\begin{equation}
 L(t)=L(t_{\rec}^-) \left\{
  \begin{array}{l l}
  \left( \frac{t}{t_{\rec}} \right)^{-\frac{2(1-s)(k-3)}{k-2}}
 \vspace{0.5cm} & \\
  L(t_{\rec}^-)
    \left( \frac{\alpha}{\alpha_L} \right)^{-\frac{(k-3)s+1}{(k-3)s+k+2}}
    \left( \frac{t}{t_{\rec}} \right)^{-\frac{2(-2(k-3)s+k-1)}{(k-3)s+k+2}} 
  \end{array} \right.
 \quad \propto \quad 
  \left\{
  \begin{array}{l l}
 t^{-0.07}& \quad,   m_{\rec}<\hat{m}
 \vspace{0.5cm} \\
     \alpha^{-0.35}
 t^{0.13} & \quad, m_{\rec}=\hat{m}\\
  \end{array} \right. .
\label{eq:lum_in_const}
\end{equation}
Equation \eqref{eq:lum_in_const} gives us the discontinuity in the luminosity 
at $t_{\rec}$ for the case where $\alpha<\alpha_L$. For $\alpha>\alpha_L$ the luminosity
is continuous throughout the whole evolution.

For completeness, we examine the times and values of $\alpha$ in which the solution given in 
\eqref{eq:mrec_const} is relevant.
In order to check the consistency of the solution above
we assume that the energy density in a small subshell behind $T=T_{\rec}$ is $aT_{\rec}^4$ and
check whether $\eta(\delta)<1$.
Taking $\epsilon(\delta)=aT_{\rec}^4$ and $\epsilon_{\rec-\out}=aT_{\rec}^4$, equation \eqref{eq:general_const_rec} reads
$\delta=\alpha r_{\rec}$.
Using this and $m_{\rec}(t)$ from equation \eqref{eq:mrec_const} in equation \eqref{eq:eta_delta},
we find
\begin{equation}
 \eta(\delta=\alpha r_{\rec})=\left\{
  \begin{array}{l l}
    \alpha^{\frac{k-4}{k+1}}
	  \left( \frac{t}{t_{\rec}} \right)
	  ^{\frac{(k-3)(2(3k-2)s-11k+14)}{(k-2)(k+1)}}
 \vspace{0.5cm} \\
    \left( \frac{\alpha}{\alpha_0} \right)^{-\frac{2(k-3)s-k+6}{(k-3)s+k+2}} 
		\left(\frac{t}{t_{\rec}} \right)
		^{\frac{7(k-3)s-11k+26}{(k-3)s+k+2}}
  \end{array} \right.
 \quad \approx \quad \left\{
  \begin{array}{l l}
    \alpha^{0.28}
	  \left( \frac{t}{t_{\rec}} \right)
	  ^{-2.31}
 & \quad   m_{\rec}<\hat{m}
 \vspace{0.5cm} \\
    \left( \frac{\alpha}{\alpha_0} \right)^{-0.52} 
		\left(\frac{t}{t_{\rec}} \right)
		^{-1.85}
  & \quad m_{\rec}=\hat{m}\\
  \end{array} \right. ,
  \label{eq:eta_delta_alpha}
\end{equation}
where expression for $\alpha_0$ is
\begin{equation}
\alpha_0= \Delta^{\frac{(3k-2)(s(k-3)+1)}{(k-3)(2(k-3)s-k+6)}}
\approx \Delta^{3.6} .
\label{eq:alpha0}
\end{equation}
From equation \eqref{eq:eta_delta_alpha} we conclude that when $m_{\rec}<\hat{m}$ then the region behind $T=T_{\rec}$ is in thermal
equilibrium.
In addition, for $\alpha>\alpha_0$, the radiation behind
$T=T_{\rec}$ is in thermal equilibrium at the onset of recombination and it keeps to be in thermal equilibrium throughout
the whole evolution.
This is because $\eta(\delta= \alpha r_{\rec})<1$ for $t>t_{\rec}$.
Hence, the solution in equation \eqref{eq:mrec_const} is consistent from the onset of recombination if $\alpha>\alpha_0$.
If $\alpha<\alpha_0$ it is consistent from $t=t_1$ defined by $\eta(\delta=\alpha r_{\rec})(t_1)=1$
\begin{equation}
t_1= \left( \frac{\alpha}{\alpha_0} \right)^{\frac{2(k-3)s-k+6}{7(k-3)s-11k+6}} t_{\rec}
\approx \left( \frac{\alpha}{\alpha_0} \right)^{-0.28} t_{\rec} .
\label{eq:t_1}
\end{equation}


Next, we provide the dynamics of the recombination shell 
under the assumption that the radiation is out of thermal equilibrium
behind $T=T_{\rec}$. 
According to the result above, we consider only the situation in which $m_{\rec}=\hat{m}$, since
since $\alpha_0<\alpha_L$.
The radiation goes out of thermal equilibrium somewhere between the internal boundary of the recombination shell 
to the point where $T=T_{\rec}$.
Let us say that this transition appears in a subshell 
located inward to the point where $T=T_{\rec}$, with $\delta > \alpha r_{\rec}$.
The temperature in this subshell is a black body temperature, $T_{\cl}$.
We can take $T_{\cl} \approx T_{\rec}$, since from this subshell out to where $T=T_{\rec}$
the radiation is  out of thermal equilibrium and therefore can barely change the temperature. On the other hand
the energy density at the external boundary is much lower than $a T_{\rec}^4$.
This is the main difference between the case in which the radiation behind $T=T_{\rec}$ is thermalized
and when it is not, see figure~\ref{fig:mrec_in_const}.
Taking $\epsilon(\delta)=a T_{\rec}^4$, the right equality in equation \eqref{eq:general_const_rec}  
and the requirement $\eta(\delta)=1$ are solved for
$m_{\rec}(t)$ and $\delta(\eta=1)$. We find the expressions 
for the dynamics of the recombination shell and the bolometric luminosity,
\begin{equation}
\hat{m}(t)=m_{\rec}(t)= \hat{m}(t_{\rec}^-)
\Delta^{-\frac{k-4}{2(k-3)s-k+6}}
\left( \frac{t}{t_{\rec}} \right)^{\frac{5(k-3)}{2(k-3)s-k+6}}
\approx  \hat{m}(t_{\rec}^-)
\Delta^{-1.35}
\left( \frac{t}{t_{\rec}} \right)^{2.63} ,
\label{eq:mrec_in_const0}
\end{equation}
\begin{equation}
L(t)=L(t_{\rec}^-)
\Delta^{-\frac{(k-4)((k-3)s+1)}{(k-3)(2(k-3)s-k+6)}}
\left( \frac{t}{t_{\rec}} \right)^{\frac{(k-3)s+2k-7}{2(k-3)s-k+6}}
\approx L(t_{\rec}^-)
\Delta^{-0.45}
\left( \frac{t}{t_{\rec}} \right)^{1.38} .
\label{eq:lum_in_const0}
\end{equation}
The luminosity and the mass of the recombination shell given in \eqref{eq:mrec_in_const0} and \eqref{eq:lum_in_const0} 
do not depend on the value of $\alpha$. This is because we assume that the outermost point in thermal equilibrium
is in a subshell characterized with $\delta(\eta=1)>\alpha r_{\rec}$.
For such subshell the optical depth and the diffusion time do not depend on the value of $\alpha$. 
The width of this subshell is 
\begin{equation}
\delta(\eta=1)= \alpha_0
\left( \frac{t}{t_{\rec}} \right)^{-\frac{-7(k-3)s+11k-26}{2(k-3)s-k+6}} r_{\rec}
\approx \alpha_0
\left( \frac{t}{t_{\rec}} \right)^{-3.52} r_{\rec} ,
\label{eq:delta1}
\end{equation}
where the value of $\alpha_0$ is given in equation \eqref{eq:alpha0}.
To conclude, for $\alpha<\alpha_0$, at the onset of recombination the radiation behind $T=T_{\rec}$ is 
out of thermal equilibrium and the dynamics of the recombination shell is given by equation \eqref{eq:mrec_in_const0}.
As time progresses, we find that $\delta(\eta=1)/r_{\rec}$ decreases and that the coupling in this shell increases.
Once $\delta(\eta=1)=\alpha r_{\rec}$ the radiation behind $T=T_{\rec}$ is thermalized and the evolution of the 
recombination shell and the luminosity is given by equations \eqref{eq:mrec_const} and \eqref{eq:lum_in_const}.
The time in which this transition occurs is $t_1$ given in equation \eqref{eq:t_1}.
For $\alpha=0$ we see that $t_1 \to \infty$ so the evolution of the luminosity is given by  \eqref{eq:lum_in_const0} for
$t>t_{\rec}$.

We note here that for small values of $\alpha$, it is possible (depending on the initial conditions) 
that the optical depth in the relevant regions is lower than 1.
In that case, our analysis is not consistent.
However, since the step function model, especially with low values of $\alpha$, is not physical, we
do not discuss here how the analysis should be modified in the case of such inconsistency. 
In addition, for large values of $\alpha$ it is possible that $m_{\rec}>\hat{m}$, as opposed to the continuous model.
As recombination reaches deeper shell the temperature at the internal boundary decreases. 
Once $T_{\rec-\in}=T_{\rec}$, or equivalently $c/v(m_{\rec})=\tau_{\rec-\out}$,
the evolution is the same as before recombination starts, only now the
opacity is $\alpha \kappa_T$ instead of $\kappa_T$. We mark with $\tilde{t}$ 
the time from which this scenario holds.





\subsection{Color Temperature} \label{sec:temp_const}

We have already seen that there are several scenarios for the 
evolution of the bolometric luminosity, depending on the initial condition when recombination starts. 
The evolution of the color temperature also shows various scenarios.
When $\alpha<\alpha_0$ during the time $t_{\rec}<t<t_1$ the outermost point in thermal equilibrium is located 
behind the point of $T=T_{\rec}$ and we have approximated that $T_{\cl} \approx T_{\rec}$. In this case
the recombination shell is also the last shell which is in thermal equilibrium, so $m_{\rec}=m_{\cl}$.
Moreover, for $\alpha=0$ the color temperature is $T_{\rec}$ throughout the whole evolution ($t_1 \to \infty$). This is simply
because $\alpha=0$ means that there are no free electrons for $T<T_{\rec}$ so there are no efficient processes which can
change the temperature external to the point where $T=T_{\rec}$.
On the other hand, when the radiation behind the point with $T=T_{\rec}$ is in thermal equilibrium, the color temperature
may drop below $T_{\rec}$. It depends whether shells external to the recombination shell generate enough photons
through free-free emission such that the radiation is in thermal equilibrium. 

We find that for any choice of $\alpha$, at the onset of recombination the external boundary 
of the recombination shell is not in thermal equilibrium, therefore for $t=t_{rec}^{+}$ the
observed temperature is $T_{\rec}$. As we continue with the evolution of $m_{\rec}$ the coupling in the external boundary
increases and once $\eta_{\rec-\out}=1$, the color shell starts to move outward
from the recombination shell. From that point, the color temperature starts to decrease and $m_{\cl}(t)<m_{\rec}(t)$. 
The color shell in that case can be
found by requiring
$\eta(m_{\cl}<m_{\rec})=1$.
The color temperature is found by using equation \eqref{eq:eps} and \eqref{eq:Tbb}.
In table \ref{table1} we add more information about the times of transitions between different types of evolution.
For example, the times $\bar{t}_1$ and $\bar{t}_2$ in which the color shell is external to the recombination shell and the temperature
drops below $T_{\rec}$, or the critical value $\alpha_1$, such that for $\alpha>\alpha_1$ the temperature drops 
below $T_{\rec}$ before the luminosity changes due to recombination.
Figures \ref{fig:lum_const_acc} and \ref{fig:temp_const_acc} show the
evolution in time of the bolometric luminosity and observed temperature for various values of $\alpha$.

\newpage

\renewcommand{\arraystretch}{2}

\begin{table}[h]   

 \begin{tabular}{|l||c|c|c|c|}

\hline
\multicolumn{5}{|c|} {$0 \le \alpha < \alpha_0$} \\
\hline
\hline
Time & $t_{\rec}<t<t_1$ & $t_1 \le t < \bar{t}_1$ & $\bar{t}_1 \le t < \tilde{t}$ & $t>\tilde{t}$ \\
\hline
Evolution & $\hat{m}=m_{\rec}=m_{\cl}$ & $\hat{m}=m_{\rec}=m_{\cl}$ & $\hat{m}=m_{\rec}>m_{\cl}$ &
$\hat{m}>m_{\rec}>m_{\cl}$\\
\hline
Luminosity  & affected &
 affected &
 affected & 
 affected \\
\hline
Observed temperature  & $T_{\rec}$ & $T_{\rec}$ & 
$<T_{\rec}$ &
$<T_{\rec}$\\
\hline
\multicolumn{5}{c} {\vspace{-0.1cm} } \\

\end{tabular} 

  \begin{tabular}{|l||c|c|c|c|}

\hline
\multicolumn{4}{|c|} {$\alpha_0 \le \alpha < \alpha_L$} \\
\hline
\hline
Time & $t_{\rec} \le t < \bar{t}_1$ & $\bar{t}_1 \le t < \tilde{t}$ & $t>\tilde{t}$ \\
\hline
Evolution & $\hat{m}=m_{\rec}=m_{\cl}$ &  $\hat{m}=m_{\rec}>m_{\cl}$ &
$\hat{m}>m_{\rec}>m_{\cl}$\\
\hline
Luminosity  & affected &
  affected  &
affected  \\
\hline
Observed temperature  & $T_{\rec}$ & $<T_{\rec}$ &
$<T_{\rec}$ \\
\hline
\multicolumn{4}{c} { \vspace{-0.1cm} } \\

\end{tabular}

\begin{tabular}{|l||c|c|c|c|}

\hline
\multicolumn{5}{|c|} {$\alpha_L \le \alpha < \alpha_1$} \\
\hline
\hline
Time & $t_{\rec} \le t < t_L$ & $t_L \le t < \bar{t}_1$ & $\bar{t}_1 \le t < \tilde{t}$ & $t>\tilde{t}$ \\
\hline
Evolution & $\hat{m}>m_{\rec}=m_{\cl}$ &  $\hat{m}=m_{\rec}=m_{\cl}$ &
 $\hat{m}=m_{\rec}>m_{\cl}$ & $m_{\rec}>\hat{m}>m_{\cl}$ \\
\hline
Luminosity  & not affected &
affected &
affected &
affected   \\
\hline
Observed temperature  & $T_{\rec}$ & $T_{\rec}$ & 
$<T_{\rec}$  &
$<T_{\rec}$ \\
\hline
\multicolumn{5}{c} { \vspace{-0.1cm} } \\

\end{tabular} 

\begin{tabular}{|l||c|c|c|c|}

\hline
\multicolumn{5}{|c|} {$\alpha_1 \le \alpha \le 1$} \\
\hline
\hline
Time & $t_{\rec} \le t < \bar{t}_2 $ & $\bar{t}_2 \le t < t_L $ & $t_L \le t < \tilde{t}$ & $t>\tilde{t}$ \\
\hline
Evolution & $\hat{m}>m_{\rec}=m_{\cl}$ &  $\hat{m}>m_{\rec}>m_{\cl}$ &
 $\hat{m}=m_{\rec}>m_{\cl}$ & $m_{\rec}>\hat{m}>m_{\cl}$ \\
\hline
Luminosity  & not affected &
not affected &
affected &
affected   \\
\hline
Observed temperature  & $T_{\rec}$ & $<T_{\rec}$& 
$<T_{\rec}$  &
$<T_{\rec}$ \\
\hline
\multicolumn{5}{c} { \vspace{-0.1cm}  } \\

\end{tabular}
\renewcommand{\arraystretch}{1.2}
\begin{tabular}{p{\textwidth}}

$\alpha_0$ - for lower values of $\alpha$ the evolution is similar to $\alpha=0$ 
at $t_{\rec}<t<t_1$. \\

$t_1$ - from that time the radiation behind the point where $T=T_{\rec}$ 
is in thermal equilibrium.\\

$\bar{t}_1$ - the time when the color temperature drops below $T_{\rec}$, 
when $\hat{m}=m_{\rec}$.\\

$\alpha_L$ - for $\alpha<\alpha_L$ the recombination shell is the luminosity shell when
recombination starts.\\
 
$t_L$ - when $\alpha>\alpha_L$ this is the time when $m_{\rec}=\hat{m}$.\\

$\alpha_1$ - for $\alpha>\alpha_1$ the temperature drops below $T_{\rec}$ while $m_{\rec}<\hat{m}$.\\

$\bar{t}_2$ - the same as $\bar{t}_1$ only for the case that $m_{\rec}<\hat{m}$ 
(relevant for $\alpha<\alpha_1$).\\

$\tilde{t}$ - the time when $T_{\rec-\in}=T_{\rec}$.\\

\end{tabular}

\caption{Summary of the analytical description for the step function model.
We indicate when the luminosity is affected by recombination and when
the temperature drops below the recombination temperature.} 

\label{table1}

\end{table}
\section{Equations for the semi-analytical solution} \label{App_num}

We take the profiles $\rho_i=\rho(m,t_i)$, $E_{\ad,i}=E_{\ad}(m,t_i)$, $r_i=r(m,t_i)$ and $v(m)$ at some
early time $t_i$ \emph{before recombination starts} and consider all the shells which already reached their coasting velocity (spherical phase).
We defined the following properties of each shell
$$
\tau_i(m)=\int_{r_i}^\infty \kappa_T \rho_i dr_i^{\prime} ,\quad
t_{d,i}(m)=\int_{r_i}^\infty \frac{3\tau_i}{c} dr_i^{\prime} ,\quad
$$
We construct the profiles at later times, $t>t_i$, if there was \emph{no recombination}.
$$
\rho(m,t)=\rho_i \left( \frac{t}{t_i} \right)^{-3} ,\quad
E_{\ad}(m,t)=E_{\ad,i} \left( \frac{t}{t_i} \right)^{-1} ,\quad
t_d(m,t)= t_{d,i} \left( \frac{t}{t_i} \right)^{-1}  ,
$$
$$
\eta(m,t,T)=\frac{7\times10^5 s}{t_{d}} 
\left( \frac{\rho}{10^{-10}\text{g cm}^{-3}} \right)^{-2}
\left( \frac{k_BT}{100 \text{eV}}				\right)^{7/2} .
$$
The equation for $\eta$ is according to equation 9 in NS10.
Then, we solve the following equations, which represent the basic equations \eqref{eq:general}, 
for $\hat{m}(t)$, $m_{\cl}(t)$ and $T_{\cl}(t)$.
\begin{subequations}
\begin{equation}
\left \{
  \begin{array}{l l}
   t_d(\hat{m},t)=t & \quad   \text{, if } \hat{m}>m_{\cl}\\
   \hat{m}=m_{\rec} & \quad   \text{, otherwise}\\
  \end{array} \right.  ,
\label{eq:num_lumshell}
\end{equation}
\begin{equation}
\frac{E_{\ad}(\hat{m},t)}{t} = 
\frac{ aT_{\cl}^4 \frac{m_{\cl}}{\rho_{\cl}} }{ x_{\ion}(\rho_{\cl},T_{\cl}) t_d(m_{\cl},t)} ,
\label{eq:num_colorshell}
\end{equation}
\begin{equation}
   \eta(m_{\cl},t,T_{\cl}) x_{\ion}(\rho_{\cl},T_{\cl})^{-3} =1  ,
\label{eq:num_colorT}%
\end{equation}%
\label{eq:num_equations}%
\end{subequations} %
where $\rho_{cl}=\rho(m_{\cl},t)$. When $x_{\ion} \neq 1$ the color shell is the recombination shell.
The luminosity is given by $L=E_{\ad}(\hat{m},t)/t$.
In \S\ref{sec:example} we present the solution for two models of the ionization fraction-
$x_{\ion}$ is given by equation \eqref{eq:alpha} and $x_{\ion}(\rho,T)$ is given
by Saha equation. 

\section{$^{56}$Ni Luminosity}	\label{app:Ni}

Here we consider the luminosity due to $^{56}$Ni decay.
We focus on the relevance of the decay process as a source of energy. We do not discuss its
effect on the opacity.   
The signal from SNe explosions is dominated by two sources of energy- the shocked medium and the radioactive decay
$^{56}\text{Ni}\rightarrow^{56}\text{Co}$. We compare between these two sources 
during the plateau.
We consider the last shell which releases its energy. The energy due to the shock is calculated in 
\S\ref{sec:scaling} above,
$$
E_0 \left( \frac{v_{\sn} \cdot t_{\sn}}{R_{\star}} \right)^{-1} .
$$
The energy from the radioactive decay is given by the adiabatic cooling of the radioactive decay energy which
is emitted from $t=0$ up to $t=t_{\sn}$. It can be approximated by
$$
\epsilon_{56} \cdot \tau_{56} \left( \frac{\tau_{56}}{t_{\sn}} \right) M f_{56}(M)  ,
$$
where $M\approx 0.5 M_{\star}$, $f_{56}$ is the fraction of $^{56}$Ni in this shell, 
$\epsilon_{56}=3.9\times10^{10}$ is the energy per unit mass per unit time 
released by the decay and $\tau_{56}=8.76$ days is the half time of the decay.   
Whether the energy emission from $^{56}$Ni decay affect the plateau or not depends on the radius of the pre-SN, the explosion energy
and the $^{56}$Ni mixing in the envelope. We take $t_{\sn}=120$ days as the end of the plateau and
assume for simplicity that $f_{56}=0.1$. 
The energy of the shocked medium is 
$\approx 10^{48} \text{erg }  E_{51}^{0.5}  R_{500} M_{15}^{0.5}$ and the energy from $^{56}$Ni decay is
$\approx 2 \times 10^{47} \text{erg }  M_{15}$. 
Hence, the plateau feature in a typical RSG explosion 
is not affected by the energy release from $^{56}$Ni.

\section{Glossary of main symbols and notations}

\begin{enumerate}
  \setlength{\itemsep}{1pt}
  \setlength{\parskip}{0pt}
  \setlength{\parsep}{0pt}

	\item $t$: time since breakout. 
	\item $r$: radius.
	\item $v$: velocity.
	\item $m(r)$: mass which lies external to the radius r.
	\item $\rho$: mass density.
	\item $E_{\ad}$: internal energy of a shell due to adiabatic expansion.
	\item $t_d$: diffusion time.
	\item $L$: bolometric luminosity.
	\item $\epsilon$: energy density.
	\item $\eta$: thermal coupling coefficient, defined in Eq. \eqref{eq:eta_def}. 
	\item $x_{\ion}$: ionization fraction.
	\item $T_{\cl}$: the color temperature.
	\item $T_{\rec}$: the temperature in which there is a significant change in the ionization fraction due to recombination.
	\item $t_{\rec}$: the time in which recombination starts in the envelope, this is when the color temperature equals $T_{\rec}$.
	\item For quantity $x$ $=m,\rho,r,v$, the subscripts $x_{\cl}$, $x_{\rec}$ and the superscript $\hat{x}$
	are for the color shell, recombination shell and luminosity shell respectively. 
	\item For quantity $x$ $=\eta,\epsilon,\tau$, the subscripts $x_{\rec-\in}$ and $x_{\rec-\out}$ 
	are for the internal boundary and external boundary of the recombination shell respectively. 
	\item $\kappa_T$: Thomson opacity.	
	\item $t_L$: the time in which recombination starts to affect the bolometric luminosity.

\end{enumerate}

	\vspace{1cm}
\bibliographystyle{apj}	
\bibliography{Recombination_Effect_on_SNe_LC}

\end{document}